\documentclass[
 reprint,
superscriptaddress,
 amsmath,amssymb,
prb,
floatfix,
]{revtex4-2}

\usepackage{graphicx}
\usepackage{dcolumn}
\usepackage{bm}
\usepackage{xcolor}
\usepackage{float}
\usepackage[T1]{fontenc}
\usepackage[colorlinks,urlcolor=blue,citecolor=blue,linkcolor=blue]{hyperref}
\usepackage{verbatim}

\usepackage{ragged2e}
\usepackage[format=plain,labelfont=bf,labelsep=space]{caption}
\DeclareCaptionJustification{mym}{\justifying}
\newcommand{\red}[1]{{\color{black}#1}}

\begin{document}

\preprint{APS/123-QED}

\title{Momentum-space non-Hermitian skin effect in an exciton-polariton system}

\author{Yow-Ming (Robin) Hu}%
\thanks{These two authors contributed equally}
\author{Mateusz~Kr\'ol}%
\thanks{These two authors contributed equally}
\affiliation{Department of Quantum Science and Technology, Research School of Physics, The Australian National University, Canberra ACT 2601, Australia}
\author{Daria A. Smirnova}%
\affiliation{Department of Electronic Materials Engineering, Research School of Physics, The Australian National University, Canberra ACT 2601, Australia}
\author{Bianca~Rae~Fabricante}%
\affiliation{Department of Quantum Science and Technology, Research School of Physics, The Australian National University, Canberra ACT 2601, Australia}
\author{Karol Winkler}%
\affiliation{Technische Physik, Wilhelm-Conrad-Röntgen-Research Center for Complex Material Systems, Universität Würzburg, Am Hubland, D-97074 Würzburg, Germany}
\author{Martin Kamp}%
\affiliation{Technische Physik, Wilhelm-Conrad-Röntgen-Research Center for Complex Material Systems, Universität Würzburg, Am Hubland, D-97074 Würzburg, Germany}
\author{Christian Schneider}%
\affiliation{Institut für Physik, Fakultät V, Carl von Ossietzky Universität Oldenburg, 26129 Oldenburg, Germany}
\author{Sven Höfling}%
\affiliation{Technische Physik, Wilhelm-Conrad-Röntgen-Research Center for Complex Material Systems, Universität Würzburg, Am Hubland, D-97074 Würzburg, Germany}
\author{Timothy C. H. Liew}%
\affiliation{Division of Physics and Applied Physics, School of Physical and Mathematical Sciences, Nanyang Technological University, Singapore 637371, Singapore.}
\author{Andrew G. Truscott}%
\affiliation{Department of Quantum Science and Technology, Research School of Physics, The Australian National University, Canberra ACT 2601, Australia}
\author{Elena A. Ostrovskaya}%
\affiliation{Department of Quantum Science and Technology, Research School of Physics, The Australian National University, Canberra ACT 2601, Australia}
\author{Eliezer Estrecho}%
\email{eliezer.estrecho@anu.edu.au}
\affiliation{Department of Quantum Science and Technology, Research School of Physics, The Australian National University, Canberra ACT 2601, Australia}

\begin{abstract}
Localization of a macroscopic number of eigenstates on a real-space boundary, known as the non-Hermitian skin effect, is one of the striking topological features emerging from non-Hermiticity. Realizing this effect typically requires periodic (lattice) systems with asymmetry of intersite coupling, which is not readily available in many physical platforms. Instead, it is meticulously engineered, e.g., in photonics, which results in complex structures requiring precise fabrication steps. Here, we propose a simpler mechanism: introducing an asymmetric, purely imaginary potential in a topologically trivial system induces momentum-space localization akin to the skin effect. We experimentally demonstrate this localization using exciton polaritons, hybrid light-matter quasi-particles, in a simple engineered `round box' trap, pumped by a laser pump offset from the trap center. The effect disappears if the pump is concentric with the trap. The localization persists and becomes stronger at higher densities of polaritons, when a non-equilibrium Bose-Einstein condensate is formed and the system becomes nonlinear. Our approach offers a new route to realizing skin effects in continuous, non-periodic systems and exploring the interplay of non-Hermiticity, topology, and nonlinearity in macroscopic quantum states.

\end{abstract}

\maketitle

\noindent\textbf{Introduction.}
Wave localization, which refers to the confinement of disturbances within a limited region, significantly affects transport and propagation in a medium. For instance, Anderson localization, caused by disorder or random potentials, can greatly reduce electrical conductivity and, in extreme scenarios, turn the material into an insulator \cite{anderson1958}. Remarkably, localization can also result in advantageous transport properties, such as topological edge states that are confined to the boundaries of a topologically ordered material. These states are topologically protected and robust against many types of perturbation \cite{kane2005,kane2005z2,bernevig2006,senthil2015}, making them attractive for lossless guiding of waves, such as electrons or photon wave packets.

In addition to a random potential and topological ordering, wave localization can also be induced by non-Hermiticity \cite{hatano1996,hatano1998,bergoltz2021,ghatak2019,ozdemir2019,lin2023}. For example, asymmetric hopping \cite{yao2018,lee2019,yokomizo2019} or synthetic field flux \cite{gou2020,liang2022,li2022} in dissipative tight-binding models can result in the localization of a macroscopic number of eigenstates on one edge, a phenomenon called the non-Hermitian skin effect (NHSE). It can occur within a single band of eigenenergies and is closely related to the sensitivity of the generally complex-valued spectrum of the model to the boundary conditions \cite{hatano1998,shen2018}. Specifically, the spectrum under periodic boundary condition (PBC) is totally different from the spectrum under open boundary condition (OBC) \cite{leykam2017,xiong2018,yao2018,lee2019,zhang2020,yokomizo2019,borgnia2020,kunst2018,leykam2017,okuma2020,li2020,lin2023,zhang2022,shen2018,kawabata2019,gou2020,liang2022}.
There are numerous proposals and demonstrations of the NHSE in different experimental platforms, such as electric circuits \cite{hofmann2020}, photonic systems \cite{weidmann2020,zhou2023,zhang2025,longhi2024,zhang2021}, magnonic systems \cite{yan2021}, mechanical systems \cite{brandenbourger2019non,ghatak2020}, periodically driven single photons~\cite{xiao2020non}, ultracold atoms \cite{liang2022,guo2022,li2020,zhao2025two,tao2025imaginary}, reflected waves \cite{franca2022}, elastic media \cite{scheibner2020,wang2024}, and exciton-polariton systems \cite{kokhanchik2023,mandal2022,mandal2020nonreciprocal,
xu2022,xu2021,comaron2020non,xu2025exciton,Liang2026}.


Recently, the concept of NHSE has been generalized from discrete, tight-binding models to continuous periodic \cite{longhi2021,yokomizo2022} and non-periodic systems \cite{guo2022,yuce2023,kokhanchik2023,hu2024}. Surprisingly, the NHSE is robust and can also occur without periodic structures \cite{guo2022,yuce2023,kokhanchik2023,hu2024, yoda2025optical}, reducing the complexity of sample fabrication. In this case, the eigenstates are localized at the edge of a non-periodic trapping potential \cite{guo2022,yuce2023,kokhanchik2023,hu2024}. The localization can be induced by an imaginary vector potential \cite{hatano1998,yuce2023,hu2024}, a combination of effective onsite loss and Rashba-Dresselhaus spin-orbit coupling \cite{kokhanchik2023} or the boundary conditions \cite{hu2024}.
Recently, a continuous NHSE has been demonstrated in real space of a two-band  exciton-polariton system ~\cite{Liang2026} by taking advantage of polariton spin-orbit coupling and circular dichroism~\cite{kokhanchik2023,mandal2020nonreciprocal}.

In contrast with previous research that focused on real-space NHSE, in this work we show that NHSE can also be induced in {\em momentum space} of a continuous, non-periodic  system.  This effect arises from an imaginary potential that plays the role of a momentum-space imaginary vector potential. We experimentally demonstrate this effect in an exciton-polariton system by creating a reconfigurable pump-induced complex-valued potential and observing the localization of the polariton distributions on one edge of momentum space. We further show that the localization is controlled by the geometry of the pump. Moreover, the localization persists above the bosonic condensation threshold, when exciton polaritons macroscopically occupy a single state and the system becomes nonlinear. In this regime, increasing the pump power, which in turn increases polariton interactions, enhances the localization.


\textbf{Theory:} \textit{Imaginary vector potential.}
Let us consider a non-Hermitian Hamiltonian describing a quantum particle of mass $m$ confined in a potential $V(\mathbf{r})$ subject to an imaginary vector potential $i\nabla {A}(\mathbf{r})$, given by
\begin{equation}\label{eq:generalH}
    H=\frac{[\mathbf{p} - i\nabla A(\mathbf{r})]^2}{2m} + V(\mathbf{r})
\end{equation}
where $ \mathbf{r}$ and $\mathbf{p}=-i\hbar\nabla$ are the position and momentum operators, respectively. Under some conditions (as shown below), eigenstates of this Hamiltonian $\psi_n( \mathbf{r};A)$ are obtained by imaginary gauge transformation of the Hermitian eigenstates $\psi_n( \mathbf{r};0)$~\cite{yuce2023}
\begin{equation}\label{eq:eigenstate}
    \psi_n( \mathbf{r};A)=e^{-A(\mathbf{r})/\hbar}\psi_n(\mathbf{r};0).
\end{equation}
If the vector potential is purely imaginary (and hence $A(\mathbf{r})$ is real-valued), this transformation exponentially modulates the Hermitian eigenstates across space, which can lead to drastic localization.


As a toy model, consider a one-dimensional (1D) harmonic oscillator in a constant imaginary vector potential, described by
\begin{equation}\label{eq:1DHO}
    H = \frac{(p_x - iA_0)^2}{2m} + \frac{m\omega^2x^2}{2}.
\end{equation}
As above, the eigenstates can be written as $\psi_n(x;A_0)=e^{-A_0 x/\hbar} \psi_n(x;0)$ where $\psi_n(x;0)$ are the Hermite-Gaussian functions, the eigenstates of the Hermitian 1D harmonic oscillator. The exponential factor $e^{-A_0 x/\hbar}$ explicitly shows localization of all eigenstates in one direction. It represents the most basic manifestation of a non-Hermitian skin effect that can occur in a continuous, non-periodic system ~\cite{yuce2023}. \red{Note that the same localization of bound states will also occur for anharmonic trapping potentials $V(x)$. In two dimensions (2D), with $A(\mathbf{r})=A_0 x$, the eigenstates $\psi_n(\mathbf{r};A_0)=e^{-A_0x}\psi_n(\mathbf{r};0)$ will be exponentially localized along the $x$-direction.}


\begin{figure*}[t]
    \centering
    \includegraphics[width=\linewidth]{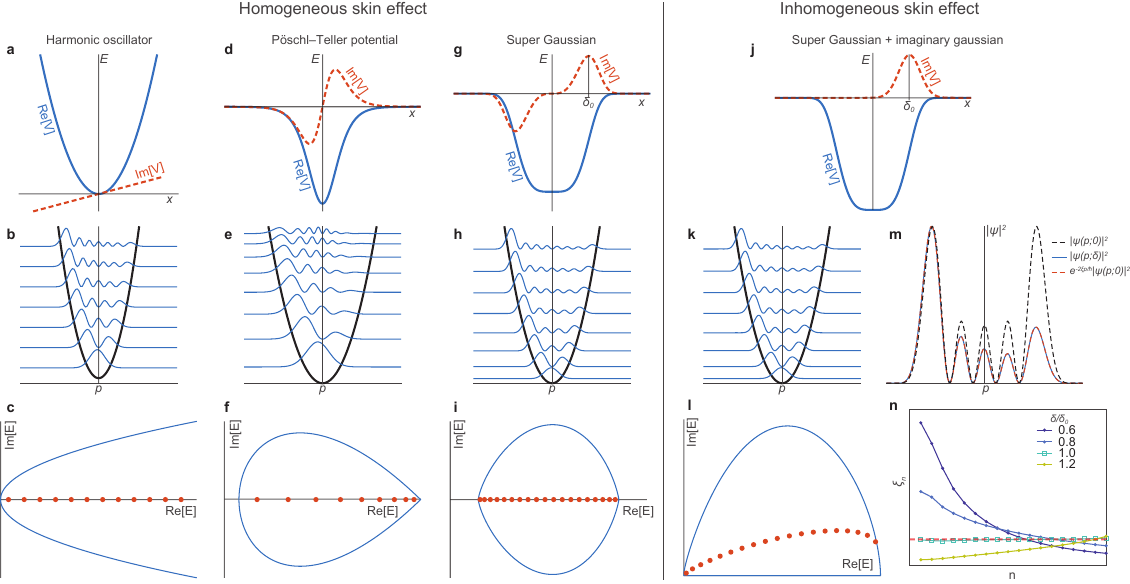}
    \caption{\red{\textbf{Continuous non-Hermitian skin effect (NHSE) in momentum space.} \textbf{a}, Harmonic oscillator potential $V(z)=z^2/2$ with imaginary shift in position $z=x+i\xi$. The imaginary part is scaled 5$\times$. \textbf{b}, Corresponding exponentially localized momentum-space eigenstates $|\psi_n(p)|^2$. Black solid line represent the kinetic energy which is the effective trap in momentum space. \textbf{c}, Corresponding spectra of eigenvalues. Solid line is the complex potential $V(x)$ corresponding to the periodic boundary conditions while the dots are the eigenstates of the potential corresponding to the open boundary condition. \textbf{d--f}, Same as \textbf{a--c} but for a P\"oshl-Teller potential $V(z)=\lambda(\lambda+1)\mathrm{sech}^2(z)/2$. \textbf{g-i}, Same as \textbf{a--c} but for an inverted super-Gaussian potential $V(z)=V_0(1-\mathrm{exp}[-(z/w)^4])$. \textbf{j--i}, Same as \textbf{g--i} but for a real-valued inverted super-Gaussian potential with an imaginary Gaussian potential $V_i\exp[-(x-\delta)^2/2\sigma^2]$. The parameters $C_0, C_1, \delta$ are chosen to match the height, width, and peak position of the positive-side imaginary part of the super-Gaussian potential in \textbf{g}. \textbf{m}, Example comparison of the momentum-space eigenstates showing that the localized-momentum eigenstate can be written as $\psi_n(p;\delta)\approx\mathrm{exp}(-\xi_n p)\psi_n(p;0)$. \textbf{n} Dependence of the decay constant $\xi_n$ on the quantum number $n$ for different imaginary Gaussian position $\delta$. The red dashed line corresponds to the half of the decay constant in \textbf{g}.}
    }
    \label{fig:fig1}
\end{figure*}

Physically realizing this real-space localization via an imaginary vector potential is not straightforward, especially in continuous systems \cite{longhi2021,guo2022,kokhanchik2023,tao2025imaginary}. In this work, we generalize the skin effect to momentum space by interchanging the roles of position and momentum, which enables a more straightforward experimental realization. In addition, we  provide further insights into more complex forms of imaginary vector potentials and their effect on the position and momentum eigenstates.

\textit{Momentum-space localization.}
\red{From another point of view, the skin effect in real space is enabled by the imaginary shift in momentum, $p\rightarrow p-iA_0$. If the same shift is induced in position (an imaginary displacement), $x\rightarrow x+i\xi$, the skin effect will occur in momentum space such that the momentum-space eigenstates $\psi_n(p;\xi)$ are exponentially localized versions of the Hermitian eigenstates $\psi_n(p;0)$, written as:
\begin{equation}\label{eq:psinp}
    \psi_n(p;\xi)=e^{-\xi p/\hbar}\psi_n(p;0).
\end{equation}
}

\red{A simple example is the 1D harmonic potential with imaginary displacement $x\rightarrow x+i\xi$, described by
\begin{equation}\label{eq:1DHOp}
    H = \frac{p_x^2}{2m} + \frac{m\omega^2(x+i\xi)^2}{2}.
\end{equation}
The eigenstates in position-space are of the form
$$\psi_n(x;\xi)\sim h_n(x+i\xi)\exp[-(x+i\xi)^2],$$
where $h_n$ are the Hermite polynomials. The momentum-space eigenstates are the Fourier transforms of $\psi_n(x;\xi)$. Using the shift theorem for Fourier transforms, the imaginary shift in $x$ manifests in a real exponential factor in $p$, which has the form given by Eq.~(\ref{eq:psinp}) where $\psi_n(p;0)$ are still the Hermite-Gauss functions. The complex potential is plotted in Fig.~\ref{fig:fig1}a and the the eigenstates $|\psi_n(p;\xi)|^2$ are plotted in Fig.~\ref{fig:fig1}b, showing signatures of the skin effect in momentum space.

The harmonic potential preserves the symmetries in Eqs.~(\ref{eq:1DHO}) and (\ref{eq:1DHOp}) by swapping the position and momentum operators, where the imaginary vector potential $A_0$ corresponds to the gradient $\xi$ of the imaginary part of the potential.
When the trapping potential is not harmonic, the imaginary potential needed to induce the skin effect will not be linear. This is relevant in the experiments presented in this work where the potential is far from harmonic.}

\red{To gain a better understanding of the effect, we look first at the analytically solvable P\"oshl-Teller potential with imaginary displacement given by $V(x;\xi)=-\lambda(\lambda+1)\mathrm{sech}^2(x-i\xi)$.
The complex potential is plotted in Fig.~\ref{fig:fig1}d, featuring an imaginary part $\mathrm{Im}[E]$ that is not linear. Despite this nonlinear function, the corresponding momentum-space eigenstates presented in Fig.~\ref{fig:fig1}e feature the skin effect, with an exponential decay consistent across all bound states following Eq.~(\ref{eq:psinp}). The explicit analytical forms of the eigenstates are detailed in the SI.}

\red{
A similar effect can be shown for an inverted super-Gaussian potential with $V(x)=V_0(1-\exp(-[(x-i\xi)/w]^{4})$. The complex potential is plotted in Fig.~\ref{fig:fig1}g. Note that $\mathrm{Im}[E]$ features two Gaussian-like peaks asymmetric to each other. There is no analytical solution but the numerically calculated eigenstates $\psi_n(p;\xi)$ are clearly localized in one direction (see Fig.~\ref{fig:fig1}h), like the previous two cases. We further show in the SI that the eigenstates follow Eq.~(\ref{eq:psinp}), as expected.

The three presented cases feature an exponential decay constant $\xi$ that is independent on the quantum number $n$, due to a well defined imaginary displacement $x\rightarrow x-i\xi$. We call these cases the ``homogeneous'' skin effect. In the following, and supported by experiments, we show that the skin effect in momentum space is robust against deviations from the exact form of the displacement, i.e. even if $V(x)\neq V(x-i\xi)$.

We demonstrate this numerically using a real super-Gaussian potential but with an eccentric imaginary Gaussian potential, described by
\begin{align}\label{eq:anharmonic}
    {\rm Re}[V]&=V_0\left(1-{\rm exp}{\left[-({x}/{w})^{4}\right]}\right), \nonumber \\
    {\rm Im}[V]&=V_i{\rm exp}[-(x-\delta)^2/(w_i^2)],
\end{align}
and plotted in Fig.~\ref{fig:fig1}j. Here, $V_i$ is the height, $\delta$ is the offset from the center, and $w_i$ is the width of the imaginary Gaussian. Interestingly, even if the imaginary displacement condition is not satisfied, the eigenstates (see Fig.~\ref{fig:fig1}k) retain the localization towards one direction in momentum, and hence the skin effect. However, the relation between the Hermitian and skin modes is modified into:
\begin{equation}\label{eq:psinpN}
    \psi_n(p;\delta,V_i)\approx e^{-\xi_n p/\hbar}\psi_n(\alpha p;\delta,0).
\end{equation}
The decay constant $\xi_n$ now depends on the quantum number $n$, making the skin effect ``inhomogeneous''. The nodes of the eigenstates also shift, as described by the scaling parameter $\alpha$, which is close to 1. An example of the excellent agreement between this approximation and the numerical simulations are presented in Fig.~\ref{fig:fig1}m. The extracted $\xi_n$ are presented in Fig.~\ref{fig:fig1}n, showing the clear dependence of $\xi_n$ on the quantum number $n$.

We have also explored how $\xi_n$ changes when the position $x=\delta$ of the imaginary Gaussian is moved, shown by different $\delta$ values in Fig.~\ref{fig:fig1}n. Surprisingly, there is a sweet spot where the $\xi_n$ is nearly constant, similar to the ``homogeneous'' skin effect induced by an imaginary displacement. This occurs when the imaginary Gaussian (see dashed orange curve in Fig.~\ref{fig:fig1}j) closely matches the positive side of the $\mathrm{Im}[E]$ of the super-Gaussian potential (see dashed orange curve in Fig.~\ref{fig:fig1}g), i.e. when $\delta=\delta_0$. Furthermore, the decay constant at this position is exactly half of the homogeneous decay constant in the super-Gaussian case (red dashed line in Fig.~\ref{fig:fig1}n). Note that when $\delta\rightarrow0$, the approximation Eq.~(\ref{eq:psinpN}) fails and the skin effect disappears since the eigenstates become symmetric both in real and momentum space.
}

\red{We can use the results above for the symmetric ($\delta=0$) and asymmetric ($\delta\neq 0)$ imaginary Gaussian to understand the 2D problem where the imaginary Gaussian is displaced to $\mathbf{r_0}=(\delta_x,0)$ from the centre of a real-valued super-Gaussian potential. We can approximately separate the problem into a symmetric (along the $y$-direction) and asymmetric (along the $x$-direction). Hence, the skin effect persists in the $k_x$ direction but disappears in the $k_y$ direction, as we have numerically shown in the SI.

Hence, the extension of Eq.~(\ref{eq:psinpN}) to this 2D problem has the form
$\psi_n(\mathbf{p};\xi)\approx e^{-\xi_n p/\hbar}\psi_n(\mathbf{p};0)$, where $\psi_n(\mathbf{p};0)$ are the 2D momentum-space eigenstates of the inverted super-Gaussian. This includes eigenstates whose nodes are oriented perpendicular to the direction to the offset of the imaginary Gaussian, further showing the robustness of the effect. Although the 2D problem is not strictly separable, our numerical results show that the skin effect in 2D is well described by the physics exhibited in the 1D problem presented in Fig.~\ref{fig:fig1}.
}

To understand the connection of the 1D  models considered here with the prototypical Hatano-Nelson model \cite{hatano1998}, the minimal discrete (tight-binding) model exhibiting the NHSE, we discretize the \red{harmonic oscillator Hamiltonian Eq.~(\ref{eq:1DHOp})} in momentum space as:
\begin{equation}\label{eq:Hktight-binding}
H_{p,n} = \epsilon_n \, |n\rangle\langle n| 
+ t_+ \, |n\rangle\langle n+1| 
+ t_- \, |n+1\rangle\langle n|,
\end{equation}
where:
\begin{align}
\epsilon_n &= \frac{n^2 (\Delta p)^2}{2m} + \frac{m\hbar^2\omega^2}{(\Delta p)^2}, \\
t_\pm &= -\frac{m\hbar^2\omega^2}{2(\Delta p)^2}\left( 1\pm \xi\Delta p \right),
\end{align}
$n$ is the index and $\Delta p$ is the lattice spacing in momentum space. Here, the imaginary vector potential $\xi$ clearly induces an asymmetric nearest-neighbor hopping ($t_+\neq t_-$), which is the defining feature of the NHSE in the Hatano-Nelson model \cite{hatano1998}. Meanwhile, the kinetic energy term plays the role of the trapping potential that only contributes to the onsite potential $\epsilon_n$.
\red{The same anisotropic hopping can be written for the other cases presented in Fig.~\ref{fig:fig1}}.

We further investigate the topological nature of the localization by examining the sensitivity of the spectrum of Eq.~(\ref{eq:Hktight-binding}) to the boundary condition. We obtain the unconfined spectra, corresponding to the periodic boundary condition (PBC), by setting $m\rightarrow\infty$ to remove the effective confinement in momentum space. Hence, the unconfined spectrum is simply the potential, i.e. $E_{\rm PBC} = V(x)$, \red{plotted as solid curves in Fig.~\ref{fig:fig1}c,f,i,l.}. The confined discrete spectrum, corresponding to the open boundary condition (OBC), \red{are the eigenvalues shown as dots in Fig.~\ref{fig:fig1}c,f,i,l. The stark difference} between the unconfined and confined spectra is analogous to the point-gap topology in 1D tight-binding models that exhibit NHSE, where the PBC spectrum encircles a finite area in the complex plane \cite{shen2018}. \red{The main difference between the homogeneous and inhomogeneous cases presented in Fig.~\ref{fig:fig1} is that when the imaginary shift condition $x\rightarrow x-i\xi$ is satisfied, the eigenvalues lie on the real axis since the system is PT symmetric~\cite{ozdemir2019}, which is not the case for the inhomogeneous skin effect. The analytic continuation of the P\"oschl-Teller potential is provided in the SI to provide further insights in the case when the PT symmetry is absent.
}

%
%
The unconfined spectrum forms a closed loop in the complex plane with a finite area and a non-zero topological winding number 
\begin{equation}
     W=\frac{1}{2\pi}\int_x \nabla_x[\arg(V(x)-E_b)] dx,
\end{equation}
which is analogous to the spectral winding corresponding to the skin modes in real space (see SI for more details), except here $x$ plays the role of $p$, and $V(x)$ plays the role of the unconfined spectrum.
\red{For the super-Gaussian with imaginary Gaussian case}, the sign of $W$ depends on the offset parameter $\delta$ in Eq. (\ref{eq:anharmonic}). When the Gaussian and the confining potential are concentric, $\delta=0$, there is no edge localization and the topological winding of the unconfined spectrum is zero (see SI for more details).



\textbf{Experiments: } \textit{Below the condensation threshold}.
To demonstrate the localization in momentum space due to an imaginary potential, we follow the geometry of the potential in Fig.~\ref{fig:fig1}j and employ exciton polaritons -- hybrid light-matter particles arising from the strong coupling of excitons (electron-hole pairs) and cavity photons in a semiconductor microcavity \cite{kavokin2017,deng2010,carusotto2013}. This system is an experimentally accessible platform for studying non-Hermitian physics, such as exceptional points, associated nontrivial dynamics, and topology~\cite{gao2015,su2021,gao2018}. It has been the subject of several theoretical investigations of non-Hermitian geometry and wave packet dynamics \cite{sedov2018,solnyshkov2021,liao2021,hu2023,hu2024qgt,hu2025}.

\begin{figure*}[t]
    \centering
    \includegraphics{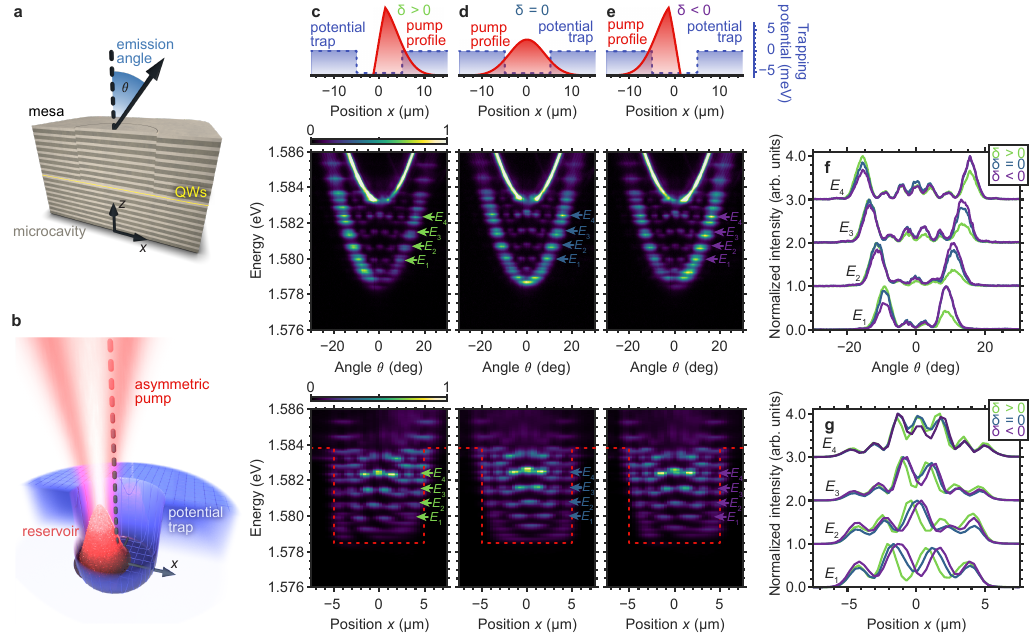}
    \caption{\textbf{Experimental momentum-space localization of polaritons in a trap with a controllable imaginary potential.} \textbf{a,} Sketch of the optical microcavity formed by distributed Bragg reflectors with embedded exciton-hosting quantum wells (QWs) and a microstructured mesa defining the trapping potential for polaritons. The angle of cavity photoluminescence emission, which determines the measured momentum ($p \propto \sin(\theta)$, see text). \textbf{b,} Schematics of the microstructured potential and the optically induced reservoir that controls the imaginary potential. \textbf{c--e,} Momentum (top) and position (bottom) resolved spectra of polaritons for three positions (see inset) of the pump laser. \red{The dashed red lines mark the trap height.} \textbf{f,} Momentum and \textbf{g,} position space profiles of four selected energy states ($E_1 = 1.5800$\,eV, $E_2 = 1.5808$\,eV, $E_3 = 1.5816$\,eV, $E_4 = 1.5824$\,eV) marked with the arrows in \textbf{c--e} (green/blue/purple for positive/zero/negative pump asymmetry in real space).}
    \label{fig:fig2}
\end{figure*}

In planar Fabry-P\'erot microcavities (see Fig.~\ref{fig:fig2}a), low-momenta exciton polaritons behave as 2D massive particles with a very small effective mass $10^{-5}$ times the rest mass of an electron. These particles have a short lifetime ($\tau\sim 10$~ps in our sample) mainly due to the photon component escaping through the imperfect mirrors of the cavity as photoluminescence (PL). The decay leads to complex energy spectra for exciton polaritons, with the imaginary part proportional to the exciton-polariton decay rate $\gamma\propto \tau^{-1}$. In our sample, the polaritons are confined in a trapping potential $V_{\rm trap}$ induced by circular mesas (see Fig.~\ref{fig:fig2}a) embedded in the sample via nanofabrication~\cite{schneider2017exciton} with a potential depth of $V_0\approx 5$~meV. These mesas can be modeled by a 2D super-Gaussian potential and the cross section is similar to the real part of the potentials shown in Figs.~\ref{fig:fig1}g,j.

To control the imaginary part of the potential, we introduce off-resonant optical pumping~\cite{gao2015,wertz2012propagation}. The pump injects carriers that form an excitonic reservoir with density $n_r (\mathbf{r})$. This reservoir scatters into low-energy polaritons at a rate $Rn_r(\mathbf{r})$ and at the same time repels them via Coulomb interactions, producing a real potential $g_r n_r(\mathbf{r})$, where $g_r$ scales with the exciton–exciton interaction strength~\cite{estrecho2019direct}. The resulting 2D complex-valued potential is:
\begin{equation}\label{eq:Vcomplex}
    V(\mathbf{r})= V_{\rm trap}(\mathbf{r}) + g_r n_r(\mathbf{r}) + i(R n_r(\mathbf{r})-\hbar\gamma). 
\end{equation}
Since $n_r(\mathbf{r})$ follows the pump intensity profile, $P(\mathbf{r})$, shaping the pump allows us to control the imaginary part of the potential. \red{Details of the 2D problem are presented in the SI}.

We control the geometry of the overall potential by moving the position of the pump with respect to the mesa trap, as illustrated in Fig.~\ref{fig:fig2}b.   The pump creates a Gaussian-shaped reservoir $n_r\propto\text{exp}[-(\mathbf{r}-\mathbf{r}_0)/2w^2]$, \red{where $\mathbf{r_0}=(\delta,0)$}. This induces a corresponding imaginary potential (see Fig.~\ref{fig:fig2}b). The reservoir $n_r$ also generates a real contribution to the overall potential (\ref{eq:Vcomplex}) but it does not directly affect the momentum-space localization because the trapped eigenstates satisfy the condition $\langle \mathbf{p} \rangle = 0$ in the Hermitian limit.

The sample emission (or photoluminescence, PL) carries information about the distributions of the eigenstates of polaritons in the trap which we can image directly and separate in energy using a spectrometer (see SI for the experimental setup). The in-plane momentum $p_x$ of polaritons in the direction of the spectrometer slit is proportional to the emission angle $\theta$ (see Fig.~\ref{fig:fig2}a) through the relation $p =2\pi \hbar \sin(\theta)/\lambda$, where $\lambda$ is the wavelength of the emitted photon (see SI for the description of experimental methods).

To simplify the analysis, we use the slit orientation (the $x$-direction) as our primary axis. We move the pump and measure the momentum (or angular distribution) and spatial distributions along the same direction. The symmetry along the $y$-direction is preserved. Hence, we can reduce the 2D problem to 1D in the following analysis. \red{This provides excellent agreement to the case presented in Fig.~\ref{fig:fig1}j--n and Eq.~(\ref{eq:anharmonic}). Note that since we only take a profile at $k_y=0$, 2D modes with nodes at $k_y=0$ are missed in the experiment. However, our detailed 2D simulations in the SI show that these modes are also localized, further showing that the skin effect persist in 2D.} 

In the first experiment, we use a 10-$\mu$m-diameter trap, which supports multiple resolvable energy levels. At low pump powers, polaritons form a thermal-like gas that occupies all available energy states, allowing us to image the trapped eigenstates $|\psi_n|^2$ as shown in the two lower panels of Fig.~\ref{fig:fig2}c--e.  The trap height is marked by the red dashed line. Eigenstates with energies below the trap height are bound states of the trap. The bright parabolic emission in momentum space arises from the untrapped polaritons outside the trap region. We then use a spatial light modulator (SLM) (see SI for details) in amplitude configuration to filter out one half of the pump. The resulting pump profiles, schematically shown in the top panel of Fig.~\ref{fig:fig2}c--e, induce imaginary potentials that are either concentric with the trap Fig.~\ref{fig:fig2}d) or are offset from the center (Fig.~\ref{fig:fig2}c,e).

In the concentric configuration shown in Fig.~\ref{fig:fig2}d, the potential $V(x)$ has both real and imaginary parts that are symmetric with respect to the trap center, resulting in symmetric position and momentum space distributions of the trapped states. The slight asymmetry arises from the tilt or wedge of the cavity length across the sample due to unavoidable variation in growth rates of epitaxial layers during fabrication \cite{Sermage2001,Steger2013}.

When the pump is offset from the center (i.e., asymmetric), it creates a strong asymmetric imaginary potential, inducing an \red{effective} imaginary vector potential in momentum space, \red{similar to the results presented in Fig.~\ref{fig:fig1}j--n}. The resulting momentum-space distributions are shown in Fig.~\ref{fig:fig2}c,e, which clearly shows the skewing of the distributions towards the $\pm k_x$-direction, depending on the asymmetry of the pump. A comparison of the profile of a chosen trapped state is shown in Fig.~\ref{fig:fig2}f, clearly showing the difference in distribution.

As expected, the real-space distribution of the trapped states only changes by a small amount, and remains symmetric, as shown by the lower panels of Fig.~\ref{fig:fig2}c--e and the cross sections in Fig.~\ref{fig:fig2}g. These small changes are primarily due to the real part of the pump-induced potential, which has a strong effect on the lowest-lying states since the induced potential is smaller than the trap height.

\begin{figure*}[htb!]
    \centering
    \includegraphics{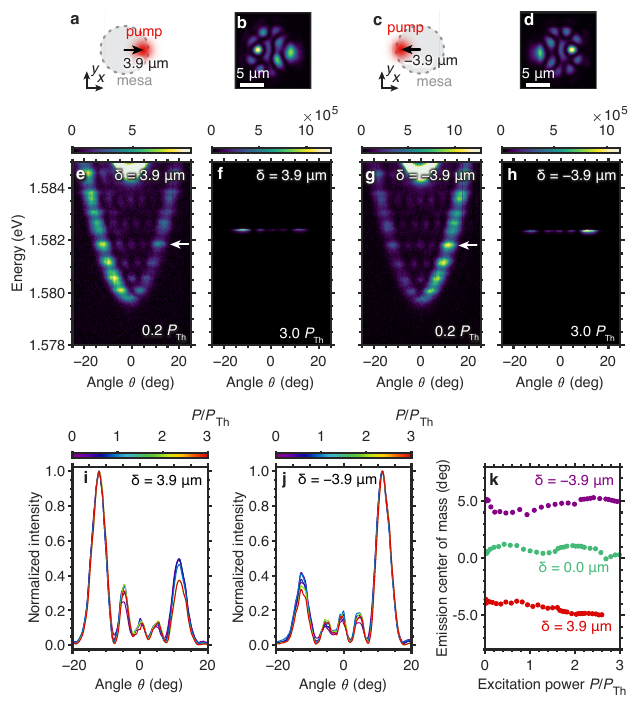}
    \caption{\textbf{Condensation of momentum-space localized polaritons.} \textbf{a,c,} Schematics of pump configuration and corresponding \textbf{b,d,} real-space distribution of the condensed mode. \textbf{e,h,} Angle-resolved PL below to above threshold for the two pump configurations. \red{The arrows in \textbf{e,g} indicate the mode where the polariton condensate form in \textbf{f,h}, which has the spatial profile in \textbf{b,d}, and momentum-space profile in \textbf{i,j}.} \textbf{i,j,} Normalized (with respect to maximum) momentum-space profile of the mode as a function of pump power. \textbf{k,} Center-of-mass momentum (or emission angle) of the condensing mode as a function of pump power.}
    \label{fig:fig3}
\end{figure*}

\textit{Localization of condensed polaritons.} We further investigate here the mode profiles of polaritons when they form a condensate, which occurs at a higher pump power, above the condensation threshold. In this second experiment, the pump shape is fixed (Gaussian with $\mathrm{FWHM} =3.1~\mu$m), but the center is moved across the trap, as schematically shown in Fig.~\ref{fig:fig3}a,c, by moving a lens in the excitation beam path (see SI for details). We then run power series measurements for different pump positions, and the results when the pump is centered $\delta = \pm 3.9~\mu$m from the center of the trap are presented in Fig.~\ref{fig:fig3}.

At pump powers below the condensation threshold $P_{\rm th}$, the momentum-space distributions (see Fig.~\ref{fig:fig3}e,g) feature occupation of multiple states skewed towards the $\pm k_x$ (or $\pm \theta$) directions, reminiscent of the non-Hermitian skin effect and similar to the case in Fig.~\ref{fig:fig2}. Note that the Gaussian pump shape in this second experiment, shown in Fig.~\ref{fig:fig3}a,c, is different to the masked Gaussian profile shown in  Fig.~\ref{fig:fig2}c,e, although they both feature the same asymmetry with respect to the trap center. This strongly suggests that the localization is robust and not sensitive to the actual shape of the imaginary potential, as long as there is a strong asymmetry of the pump.

At pump powers above $P_{\rm th}$, polaritons macroscopically occupy a single or a few modes of the trap, forming a non-equilibrium polariton condensate \cite{Sanvitto2009,Maragkou2010}. This results in a dramatic change of the spectral distribution above threshold (see Fig.~\ref{fig:fig3}f,h). The condensation is accompanied by a nonlinear increase of the PL intensity, blueshift, and linewidth narrowing of the condensed mode (see Fig.~S6 in SI).

In the two pump configurations presented in Fig.~\ref{fig:fig3}f,h, the five-lobed (in the shown cross-section) mode dominates the spectrum at high pump powers. When the condensate forms in a single mode at sufficiently high pump powers, the full 2D real-space distribution of the condensed mode can be directly measured, as shown in Fig.~\ref{fig:fig3}b,d. Note that in other pump positions, the condensate forms in more than one mode or different set of modes so we choose the pump configurations presented in Fig.~\ref{fig:fig3} for which the condensate forms in a single and similar mode with opposite localization directions (see SI for the results obtained for other pump positions).

Interestingly, the momentum-space profile shows enhancement of localization with increasing pump power above threshold. For example, the intensity of the rightmost peak in Fig.~\ref{fig:fig3}i decreases with increasing pump power relative to the main peak, suggesting an increase in localization. The same effect is observed for the opposite pump configuration (see Fig.~\ref{fig:fig3}j)

To quantify the effect, we calculate the momentum-space  center of mass (COM) of the condensed mode as a function of pump power with the results shown in Fig.~\ref{fig:fig3}k. The trends clearly show that the COM moves further away from the center (hence increasing localization) with increasing pump power above the threshold. Since the condensate density increases with pump power, the results suggest that polariton interactions favor localization enhancement.

\begin{figure*}
    \centering
    \includegraphics{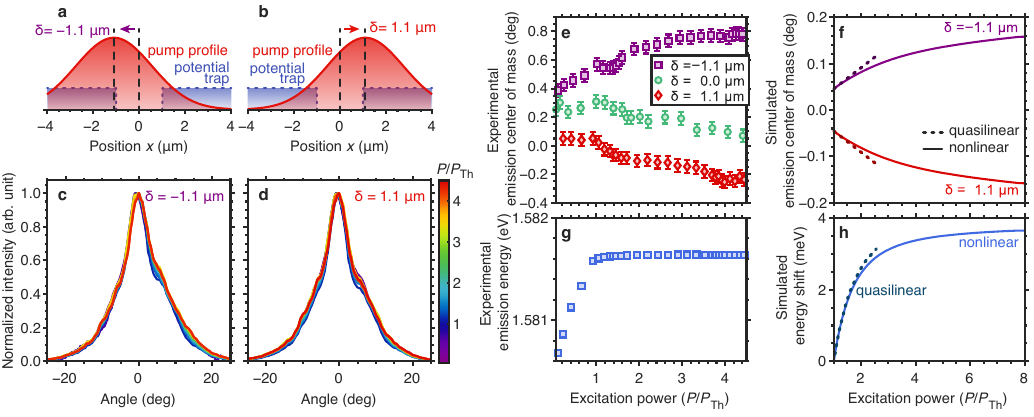}
    \caption{\textbf{Ground-state condensation of momentum-space localized polaritons in a tight trap.} \textbf{a,b,} Two pumping configurations. \textbf{c,d,} Ground-state angle-resolved distribution and \textbf{e,} experimental momentum-space COM as a function of pump power. \textbf{f,}~Simulated COM as a function of pump power. \textbf{g,h,} Experimental \textbf{g} and simulated \textbf{h} condensate energy as a function of pump power.}
    \label{fig:fig4}
\end{figure*}

To probe this interaction-induced enhancement of localization further, we use a small trap ($D=2~\mu$m) to confine condensed polaritons in a small area, resulting in large densities. In this configuration, the condensate forms in the ground state. The Gaussian pump spot ($\text{FWHM} = 3.9$\,$\mu$m) is bigger than the trap but we can still reliably create an offset between the pump and the trap center with a clear difference in gradient of the pump (or imaginary potential) across the trap, as shown in Fig.~\ref{fig:fig4}a,b. The momentum-space profile of the ground state condensate for the two pump configurations are shown in Fig.~\ref{fig:fig4}c,d as a function of pump powers. The momentum-space COM in this trap (see Fig.~\ref{fig:fig4}e) shows a behaviour that is very similar to that in a larger trap with a high-order mode condensation in Fig.~\ref{fig:fig3}i. There is a very clear trend of the momentum COM moving away from the center with increasing pump power above threshold (i.e., increasing condensate density). This further supports our conclusion that polariton interaction enhances the momentum-space localization.

\textbf{Numerical Simulations.} 
We model the behaviour of the polariton wavefunction within the framework of a driven-dissipative Gross-Pitaevskii equation coupled to a rate equation for pump-induced reservoir density, $n_R$ (see SI). In the resulting non-Hermitian nonlinear system, increasing the pump power affects (i) the reservoir density, leading to a slight skew of the real potential from the symmetric trap shape; (ii) the imaginary potential; and (iii) the polariton density, which drives the nonlinear interactions. To reduce computational time, we perform simulations in a one-dimensional setting.

In contrast to the linear PT-symmetric case with a symmetric real part and an antisymmetric imaginary  potential, where the spectrum of bound modes is entirely real (as exemplified in the conceptual Fig.~\ref{fig:fig1}a,b,d), the spectrum with an asymmetric imaginary potential is intrinsically complex-valued (such as in Fig.~\ref{fig:fig1}c,e,f). Below the condensation threshold, i.e. for $P/P_{\rm th}<1$ in Fig.~\ref{fig:fig4}f,h, the confined eigenstates are weakly dissipating and have a finite lifetime, with the imaginary parts of the eigenenergies $\mathrm{Im}(E)<0$. Increasing the pump power can bring the modes above the $\mathrm{Im}(E)=0$ line into the gain region, $\mathrm{Im}(E)>0$. The power threshold for this transition depends on the specific pumping profile and intensity. Above the threshold, it is then the nonlinearity that balances gain and loss, enabling the stationary state. 

The quasilinear solution, valid when the condensate density is still low and the terms with $|\psi|^2$ can be neglected, applies near the threshold. It can be obtained by first solving the linear eigenvalue problem with a complex-valued potential and then obtaining the mode amplitude that produces the nonlinear potential correction compensating the positive imaginary part of a given eigenenergy. In this way, it is possible to reconstruct both the shift in the mean momentum $\langle p \rangle$ and the power-dependent dispersion (momentum-resolved energy) slightly above the threshold, as shown in Fig.~\ref{fig:fig4}f,h (dashed lines). 

Once the condensate density grows, nonlinear effects become more pronounced. We then solve the stationary nonlinear problem numerically using Newton’s method to obtain the behaviour of the condensate energy. The calculated energy blueshift saturates at higher powers (see Fig.~\ref{fig:fig4}h), similar to the behaviour observed in the experiment (see Fig.~\ref{fig:fig4}g).

As the power increases, the mean momentum $\langle p \rangle$ continues to shift but gradually slows down, as seen in experiments (see Fig.~\ref{fig:fig4}e). We reproduce this behaviour by numerical modeling for a narrow trap with the ground state condensation in Fig.~\ref{fig:fig4}f and for a wider trap with the condensate in the third-order mode (see Supplementary Information). The asymmetric pump centered on one side of the trap pushes the momentum-space distribution in the opposite direction. 
Similar to the Hermitian nonlinear Schr\"odinger equation, repulsive interactions increase the condensate size in real space while decreasing it in momentum space. This results in a stronger localization factor $|\langle p \rangle/\Delta p|$, where $\Delta p$ is the momentum-space spread, compared to localization due to the linear skin effect. Note the distinct roles of nonlinearity in these two regimes: while real-space NHSE localisation is suppressed by the spreading induced by repulsive interactions~\cite{tao2025imaginary}, momentum-space NHSE localisation is enhanced because the same repulsive interactions cause the condensate to shrink in momentum space.

\textbf{Discussion.}
In this work, we have shown that a localization of eigenstates via the non-Hermitian skin effect can be induced in momentum space using an appropriate imaginary potential. The effect is analogous to the localization in real space arising from an imaginary vector potential.  We have linked the momentum-space localization to the point-gap topology of the complex eigenvalues arising from the sensitivity of the spectrum to boundary conditions. 

We have demonstrated the skin effect in momentum space experimentally in an exciton-polariton system. By controlling the imaginary part of the potential using the optical pump geometry, we have shown control over the momentum-space localization of the spatially confined modes.

We further demonstrated, both experimentally and through numerical modelling, that the localization persists and is enhanced above the bosonic condensation threshold, when polaritons form a macroscopic quantum state and display a bright, coherent light emission (polariton lasing). The center-of-mass momentum of the condensate can be tuned by changing the asymmetry of the pump, enabling dynamical and ultrafast tuning of the polariton lasing emission angle, time-limited only by the response of the excitonic reservoir. This provides a new mechanism  for changing the angle of emission of polariton laser, which can be useful for ultrafast laser beam steering applications.




It is important to note that the localization in our experiment is weak, because the imaginary potential is relatively weak. It will be interesting to investigate how strong imaginary potentials (and stronger localization) can be induced, perhaps in other experimental platforms or other geometries~\cite{zhang2022imaging}. 

Finally, our work paves the way for experimentally probing the intersection of non-Hermitian, topological, and nonlinear physics~\cite{bergoltz2021, smirnova2020nonlinear}. 




\textbf{Methods.} The measurement setup is shown in SI. The sample in a cryostat at a temperature of 4.5~K was excited with a continuous wave Ti:sapphire laser set to  1.715~eV. To avoid sample heating, the laser power was modulated with an AOM with a duty cycle of 0.5\% and a frequency of 1~kHz. The asymmetry of the excitation beam was created either with a spatial light modulator or by translation of a convex lens on the laser path. The detailed description of the excitation setup is provided in SI. In both configurations, the laser beam is focused on the sample with a 50$\times$ microscope objective. Angle-resolved spectra were obtained by imaging the back focal plane of the objective on an entrance slit of a spectrometer equipped with an EMCCD camera. 

\textbf{Data availability.} The data used in this study are available in Zenodo under the accession code \url{https://doi.org/10.5281/zenodo.22113976}.

\textbf{Code availability.} The custom simulation codes used to generate the findings of this study are available from the corresponding author upon reasonable request.

\textbf{Acknowledgements.} We thank Lev Smirnov for their insightful feedback and valuable discussions. The project was supported by the Australian Research Council (ARC) through the Centre of Excellence Grant CE170100039, the Discovery Early Career Researcher Award DE220100712, and the Future Fellowship FT230100058, and by the Australian Government Research Training Program (RTP) Scholarship.

\textbf{Funding declaration.} Y.M.R.H discloses support for the research of this work from the Australian Government Research Training Program (RTP) Scholarship. Y.M.R.H, M.K., B.R.F. E.A.O. and E.E. disclose support from the Australian Research Council (CE170100039). E.E. discloses support from the Australian Research Council (DE220100712). D.A.S discloses support from Australian Research Council (FT230100058)

\textbf{Competing Interests.} The authors declare no competing interests.

\bibliography{ref_polariton}

\end{document}


\preprint{APS/123-QED}

\title{Supplementary information for Momentum-space non-Hermitian skin effect in an exciton-polariton system}

\author{Yow-Ming (Robin) Hu}%
\thanks{These two authors contributed equally}
\author{Mateusz~Kr\'ol}%
\thanks{These two authors contributed equally}
\affiliation{Department of Quantum Science and Technology, Research School of Physics, The Australian National University, Canberra ACT 2601, Australia}
\author{Daria Smirnova}%
\affiliation{Department of Electronic Materials Engineering, Research School of Physics, The Australian National University, Canberra ACT 2601, Australia}
\author{Bianca~Rae~Fabricante}%
\affiliation{Department of Quantum Science and Technology, Research School of Physics, The Australian National University, Canberra ACT 2601, Australia}
\author{Karol Winkler}%
\affiliation{Technische Physik, Wilhelm-Conrad-Röntgen-Research Center for Complex Material Systems, Universität Würzburg, Am Hubland, D-97074 Würzburg, Germany}
\author{Martin Kamp}%
\affiliation{Technische Physik, Wilhelm-Conrad-Röntgen-Research Center for Complex Material Systems, Universität Würzburg, Am Hubland, D-97074 Würzburg, Germany}
\author{Christian Schneider}%
\affiliation{Institut für Physik, Fakultät V, Carl von Ossietzky Universität Oldenburg, 26129 Oldenburg, Germany}
\author{Sven Höfling}%
\affiliation{Technische Physik, Wilhelm-Conrad-Röntgen-Research Center for Complex Material Systems, Universität Würzburg, Am Hubland, D-97074 Würzburg, Germany}
\author{Timothy~C.H.~Liew}%
\affiliation{Division of Physics and Applied Physics, School of Physical and Mathematical Sciences, Nanyang Technological University, Singapore 637371, Singapore.}
\author{Andrew G. Truscott}%
\affiliation{Department of Quantum Science and Technology, Research School of Physics, The Australian National University, Canberra ACT 2601, Australia}
\author{Elena A. Ostrovskaya}%
\affiliation{Department of Quantum Science and Technology, Research School of Physics, The Australian National University, Canberra ACT 2601, Australia}
\author{Eliezer Estrecho}%
\email{eliezer.estrecho@anu.edu.au}
\affiliation{Department of Quantum Science and Technology, Research School of Physics, The Australian National University, Canberra ACT 2601, Australia}

\maketitle

\section{Topological Origin of Skin Effects in a Continuum}
\label{sec:TopologicalOrigin}
The topological origin of the non-Hermitian skin effect (NHSE) is established through the non-trivial spectral winding and the sensitivity of the spectra to the boundary conditions \cite{zhang2020,okuma2020,yao2018}. In prototypical tight-binding models, the NHSE is accompanied by the sharp difference between spectra under the periodic boundary condition (PBC) and the open-boundary condition (OBC). The OBC spectrum is enclosed by the PBC spectrum that have a non-trivial spectral winding in the complex plane \cite{zhang2020,okuma2020,kawabata2019,shen2018}.

\begin{figure}[b]
    \centering
    \includegraphics[width=0.45\textwidth]{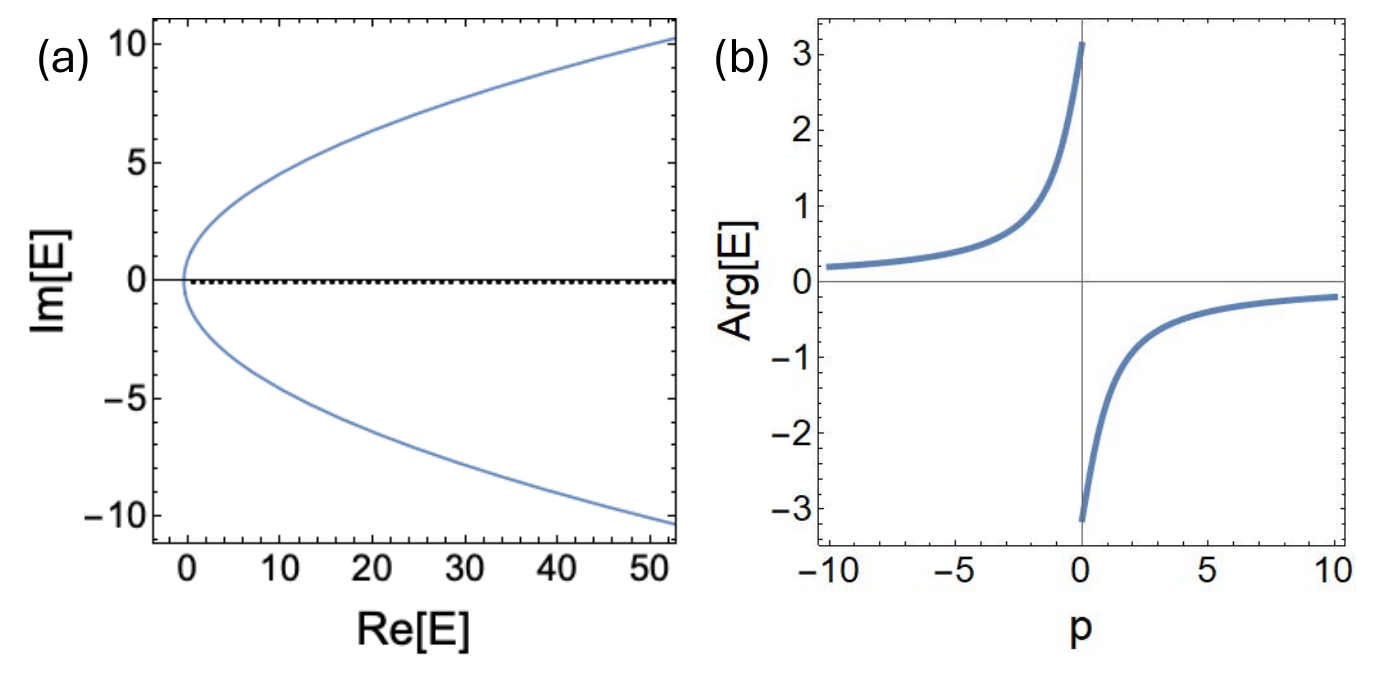}
    \caption{(a) The confined (blue line) and unconfined spectrum (black dots) of the Hamiltonian (\ref{eq:EV0_RS}) with harmonic potential defined by Eq.~(\ref{eq:harmonic_V}). (b) The complex angle of the free spectrum, where the discontinuous jump at $p=0$ results in non-zero spectral winding.}
    \label{fig:fig0}
\end{figure}

In the continuous version of NHSE, we replace PBC with the unconfined condition, i.e. there is no confining potential, and use the momentum range of $p\in(-\infty,\infty)$ instead of a periodic one. The confinement due to the potential $V(x)$ plays the role of OBC \cite{kokhanchik2023,guo2022}. The unconfined spectrum is simply
\begin{equation}
    E_{V=0} = \frac{[ p - iA_0]^2}{2m}, \quad p\in(-\infty,\infty),    
    \label{eq:EV0_RS}
\end{equation}
a parabola in the complex plane, while the trapped spectrum remains unchanged by the imaginary gauge transformation $E=E_{A_0=0}$ \cite{longhi2021,yuce2023}. For an illustrative example, consider a quantum harmonic oscillator with
\begin{equation}\label{eq:harmonic_V}
   V(x)=\frac{1}{2}m\omega x^2,
\end{equation}
where $\omega$ is the trapping frequency, the harmonically trapped spectrum is
\begin{equation}\label{eq:En_RS}
    E_n=(n+1/2)\hbar\omega, \quad n\in\mathbb{N}
\end{equation}
which is purely real-valued. The two spectra are plotted in the complex plane in Fig.~\ref{fig:fig0}(a), showing the stark difference between the two, establishing the sensitivity to the boundary condition.

Next, we can calculate the spectral winding of the unconfined spectrum around some base energy $E_b$ for the range $k=(-\infty,\infty)$ following the topological index defined for tight-binding models \cite{shen2018,zhang2020,kawabata2019} as follows:
\begin{equation}\label{eq:winding}
    w=\frac{1}{2\pi}\int_k \nabla_k[\arg(E_{V=0}-E_b)]\cdot dk = \mathrm{sgn}(A_0).
\end{equation}
The nontrivial winding depends on the sign of $A_0$, which determines the direction of localization. This is equivalent to the point-gap topology in tight-binding models, which generalizes the bulk-boundary correspondence to non-Hermitian systems, and in this case extends to continuous systems.

\section{Harmonic potential with imaginary displacement}
We present here the analytical forms of the solutions to the harmonic potential with imaginary displacement given by the Hamiltonian:
\begin{equation}
    H=-\frac{\hbar^2 }{2m}\frac{\partial^2}{\partial x^2} + \frac{1}{2}m\omega^2(x+i\xi)^2,
\end{equation}
where $\omega$ is the trap frequency and $\xi$ is the imaginary displacement, or the imaginary vector potential in momentum space.

The eigenvalues of this Hamiltonian are the same as in the Hermitian case with $E_n=(n+\frac{1}{2})\hbar\omega$. The corresponding eigenstates are
\begin{equation}
    \psi_n(x;\xi) \propto e^{-\frac{m\omega (x+i\xi)^2}{2\hbar}} H_n\left( \sqrt{\frac{m\omega}{\hbar}} (x+i\xi)^2 \right)
\end{equation}
where $H_n(x+i\xi)$ are the $n$-th Hermite polynomials. For example, the ground state ($n=0$ and $H_0=1$) eigenstate simplifies to:
$$
\psi_0(x;\xi) \propto  e^{-i\frac{m\omega x\xi}{\hbar}} e^{\frac{m\omega \xi^2}{2\hbar}} e^{-\frac{m\omega x^2}{2\hbar}}.
$$
The non-Hermitian eigenstate now includes a plane wave factor $e^{-i\frac{m\omega x\xi}{\hbar}}$ and an amplification term $e^{\frac{m\omega \xi^2}{2\hbar}}$, which can be absorbed by the normalisation.

To find the eigenstates in momentum space, we take the Fourier transform
\begin{equation}
    \psi_n(p;\xi)=\frac{1}{\sqrt{2\pi\hbar}} \int^\infty_{-\infty}\psi_n(x;\xi)e^{-\frac{ipx}{\hbar}}dx.
\end{equation}
Applying a change of variables $z=x+i\xi$, such that $dz=dx$ and the complex $z$ variable runs along a horizontal line in the complex plane shifted upwards by $i\xi$, the integral yields:
\begin{align}
    \psi_n(p;\xi) &= \frac{1}{\sqrt{2\pi\hbar}} \int^{\infty}_{-\infty}\psi_n(z)e^{-\frac{ipz}{\hbar}} e^{-\frac{\xi p}{\hbar}} dz \\
    \psi_n(p;\xi) &= e^{-\frac{\xi p} {\hbar}} \left[ \frac{1}{\sqrt{2\pi\hbar}} \int^{\infty}_{-\infty}\psi_n(x;0)e^{-\frac{ipx}{\hbar}} dx \right] \\
    \psi_n(p;\xi) &= e^{-\frac{\xi p}{\hbar}} \psi_n(p;0).
\end{align}
On the second line, the integration path is shifted back to the real axis, and the exponential factor $e^{-\frac{\xi p}{\hbar}}$ is factored out of the integral. The result is equivalent to the shift theorem for Fourier transforms. The integral simply equates to the Fourier transform of the Hermitian eigenstate, which has the form:
$$
\psi_n(p;0) \propto (-i)^n e^{-\frac{p^2}{2m\hbar\omega}} H_n\left( \frac{p}{\sqrt{m\hbar\omega}}\right).
$$
The results above explicitly show that the momentum-space eigenstates are simply the Hermitian eigenstate enveloped by an exponential decay, which is the manifestation of the non-Hermitian skin effect.

The corresponding analysis of the complex energy is very similar to Section~\ref{sec:TopologicalOrigin} where the equivalent OBC is given by the same equation for the energy eigenvalues but the PBC equivalent PBC is simply given by the potential $V(x)=m\omega^2(x+i\xi)^2/2)$, which forms a parabola in the complex plane (see Fig.~1c of the main text).

\section{The Analytically Continued P\"oschl-Teller Potential}

The P\"oschl-Teller (PT) potential is a cornerstone of exactly solvable quantum mechanics, typically defined in the Hermitian limit by $V(x) = -\lambda(\lambda+1)\text{sech}^2(x)$. To explore non-Hermitian extension, we consider the analytical continuation of the spatial coordinate into the complex plane via the transformation $x \to x + i\xi$, where $\xi \in \mathbb{R}$ represents a complex shift. This deformation is equivalent to a complex gauge transformation.

\begin{figure}[t]
    \centering
    \includegraphics[width=0.3\textwidth]{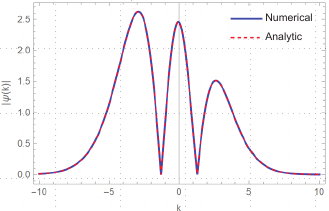}
    \caption{Example comparison of the numerical and analytic Eigenstate ($n=2$) of the PT potential for $\lambda=5$, and $\xi=0.6$.}
    \label{fig:PT_fig1}
\end{figure}

The steady-state wavefunction satisfies the stationary Schrödinger equation (assuming $\hbar= m = 1$) 
\begin{equation}
-\frac{1}{2}\frac{d^2 \psi_n}{dx^2} + V(z)\psi_n = E_n \psi_n,
\end{equation}
where derivatives are taken with respect to the real variable \(x\), while \(z = x + i\xi\). The potential is given by
\begin{equation}
V(z) = -\frac{1}{2}\lambda(\lambda+1)\,\mathrm{sech}^2(z),
\end{equation}
where \(\lambda > 0\) controls the depth of the well. $V(z)$ can be explicitly separated into the real and imaginary parts,
\begin{equation}
V(z)=V_r(x)+ i V_i(x),
\end{equation}
related through the Cauchy--Riemann conditions. 
The decomposition yields
\begin{align}
V_r(x) &=
-\frac{\lambda(\lambda+1)\left[1+\cos(2\xi)\cosh(2x)\right]}
{\left[\cos(2\xi)+\cosh(2x)\right]^2},
\\[1ex]
V_i(x) &=
\frac{\lambda(\lambda+1)\sin(2\xi)\sinh(2x)}
{\left[\cos(2\xi)+\cosh(2x)\right]^2}.
\end{align}
As expected from the analyticity of the shifted potential, the real part
$V_r(x)$ is even under $x\rightarrow -x$, whereas the imaginary part
$V_i(x)$ is odd. In the limit of a small complex shift, $\xi\rightarrow 0$, the imaginary
part is proportional to the spatial derivative of the real potential,
\begin{align}
V_i(x)
=&
\xi\,\lambda(\lambda+1)\mathrm{sech}^2(x)\tanh(x)
+\mathcal{O}(\xi^3) \\
=& \xi\,\frac{dV_r(x)\big|_{\xi=0}}{dx}
+\mathcal{O}(\xi^3).
\end{align}

The corresponding bound-state energies are
\begin{equation}
E_n = -\frac{1}{2}(\lambda - n)^2, \qquad n = 0,1,2,\dots < \lambda.
\end{equation}

The stationary states of the system, indexed by the 
quantum number $n$, are expressed in terms of the Jacobi polynomials $P_n^{(\alpha, \beta)}(z)$ of the the complex argument $x + i\xi$. 
The exact eigenfunctions analytically continued to complex \(z\) take the form
\begin{equation}
\psi_n(x; \xi) = \mathcal{N}_n \, \text{sech}^\alpha(x + i\xi) P_n^{(\alpha, \alpha)}(\tanh(x + i\xi))
\end{equation}
where shape parameter $\alpha = \lambda - n$ (with $\lambda > n$) determines the decay rate and $\mathcal{N}_n$ is the normalization constant. 

\textit{Fourier Transformation and Special Function Representation:}
To understand the momentum-space representation of these states, we evaluate the Fourier transform:
\begin{equation}
\Psi_n(k; \xi) = \int_{-\infty}^{\infty} e^{-ikx} \psi_n(x; \xi) dx
\end{equation}
Using the shift theorem for Fourier transforms, the imaginary displacement in $x$ manifests as a real exponential factor in $k$. Specifically, if $\Psi_n(k; 0)$ is the transform of the unshifted state, then:
\begin{equation}
\Psi_n(k; \xi) = e^{-\xi k} \Psi_n(k; 0)
\end{equation}
The analytical evaluation of the integral for arbitrary $n$ requires the use of generalized hypergeometric functions. 
Using the identity
\begin{equation}
P_n^{(\alpha,\alpha)}(u)
=
\frac{(-1)^n}{2^n n!}(1-u^2)^{-\alpha}
\frac{d^n}{du^n}\left[(1-u^2)^{n+\alpha}\right],
\end{equation}
with the substitution \(u=\tanh x\), and (after integration by parts \(n\) times) the Euler Beta function identity linking it to the Gamma functions,  
the 
Fourier transform 
of the $n$-th state is found to be proportional to:
\begin{equation}
\Psi_n(k; \xi) \propto e^{-\xi k} \Gamma\left(\frac{\alpha-ik}{2}\right) \Gamma\left(\frac{\alpha+ik}{2}\right) {}_3F_2\left(-n, n+2\alpha+1, \frac{\alpha+ik}{2}; \alpha+1, \alpha; 1\right)
\end{equation}
The ${}_3F_2$ term is a terminating series (since $-n$ is a negative integer), ensuring that the solution remains a finite combination of $\Gamma$ functions. For the ground state ($n=0$), the ${}_3F_2$ reduces to unity, and the profile follows the classic $\text{sech}(k)$ type decay of the $\Gamma$ product. For $n \geq 1$, the polynomial structure in $k$ introduces nodes in the momentum distribution, which are subsequently skewed by the $e^{-\xi k}$ factor. The excellent agreement of the numerical and analytic eigenstate in momentum space is presented in Fig.~\ref{fig:PT_fig1} for $n=2$.

\textit{Momentum Localization and the Non-Hermitian Skin Effect}:
The most striking consequence of the complex shift $\xi$ is the exponential ``tilt'' of the momentum-space wavefunction. In the Hermitian limit ($\xi = 0$), $|\Psi_n(k)|^2$ is symmetric about $k=0$. However, for $\xi \neq 0$, the factor $e^{-\xi k}$ induces a preferential localization in $k$-space. For $\xi > 0$, low-momentum components ($k \ll 0$) are exponentially amplified, while positive momentum components are suppressed. The center of mass of the wavepacket in $k$-space shifts toward the direction of $(-\xi)$, effectively creating a momentum-space skin effect.

In our P\"oschl-Teller model, the analytical continuation $x \to x + i\xi$ acts as this imaginary gauge field. The exponential factor $e^{-\xi k}$ is the Fourier-space manifestation of the skin effect.

Note that $\xi$ may be spatially inhomogeneous, which can lead to a more complicated kernel in the Fourier transform. 

\begin{figure*}[t]
    \centering
    \includegraphics[width=0.8\textwidth]{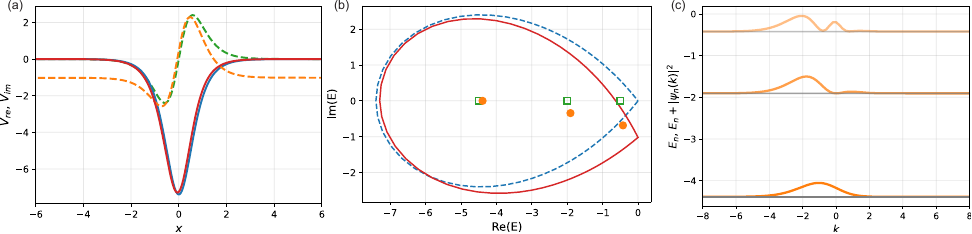}
\caption{
(a) Analytically continued P\"oschl--Teller potential without (solid red: real part; dashed green: imaginary part) and with asymmetric deformation \eqref{eq:PT_asym} and uniform decay factor $\varkappa$ (solid blue: real part; dashed orange: imaginary part). 
Parameters: $\lambda = 3$, $\xi = 0.45$, $\beta = -0.2$, $\alpha = 0$, $x_0 = 0$, and $\kappa = 1.02$.
(b) Complex energy for the case with (solid line, circles) and without (dashed line, squares) asymmetric deformation. 
(c) Eigenstates in momentum space exhibiting the skin effect for the case with asymmetric deformation.
}
    \label{fig:PT_fig2}
\end{figure*}

\textit{Complex energy mapping:}
The discrete eigenvalues $E_n$ in the complex energy plane reside within the area enclosed by the 
trajectory 
parametrized by the spatial coordinate $x \in \mathbb{R}$,
\begin{equation}
E(x) = V_r(x) + iV_i(x)\:.
\end{equation}
As $x$ sweeps from $-\infty$ to $+\infty$,
the energy $E(x)$ traces an asymptotically {closed loop}. The discrete eigenvalues $E_n$ represent the ``finite'' problem for confining potential $\frac{1}{2}k^2$.

In the purely antisymmetric case, the discrete levels $E_n$ remain real and appear as points on the real axis (in the $\mathcal{PT}$-symmetric phase). 

To break this rigid antisymmetry  of the imaginary part, 
we introduce a prefactor $F(z)$:
\begin{equation}
V(z) = -\frac{1}{2}\lambda(\lambda+1)\,\mathrm{sech}^2(z)\,F(z),
\end{equation}
admitting a complex Gaussian envelope,
\begin{equation}
F(z) = \exp\!\left[-\alpha (z - x_0)^2\right]\exp{[i\beta]},
\end{equation}
with $\alpha \ge 0$ and real shift $x_0$.
Substituting $z = x + i\xi$, the potential becomes
\begin{equation} \label{eq:PT_asym}
V(x) = -\frac{1}{2}\lambda(\lambda+1)\,\mathrm{sech}^2(x + i\xi)\,
\exp\!\left[-\alpha (x + i\xi - x_0)^2\right]\exp{[i\beta]}.
\end{equation}
This introduces an additional complex phase and amplitude modulation that is not even in $x$. As a result,  
the strict parity relation $V_i(-x) = -V_i(x)$ is no longer satisfied, however, $V_r(x)$ remains approximately symmetric, depending on the strength of the prefactor. The complex potential is plotted in Fig.~\ref{fig:PT_fig2}(a).

By perturbatively breaking $\mathcal{PT}$ symmetry, the eigenvalues $E_n$ move off the real axis [see Fig.~\ref{fig:PT_fig2}(b)], while the effect on the Fourier transform $|\Psi_n(k;\xi)|^2$ of the $n$-th bound state remains small [see Fig.~\ref{fig:PT_fig2}(c)].

If loss is already present in the system, for example via $V(x) \rightarrow V(x) - i\varkappa$, the gain provided by the asymmetric complex potential can ultimately compensate for the loss at a given level, restoring an effectively infinite lifetime (see the example in Fig.~\ref{fig:PT_fig2}(b) for $n=0$), while all other states remain lossy.

\section{Complex-Valued Pump Induced Potential}

\begin{figure*}[b]
    \centering
    \includegraphics[width=0.7\textwidth]{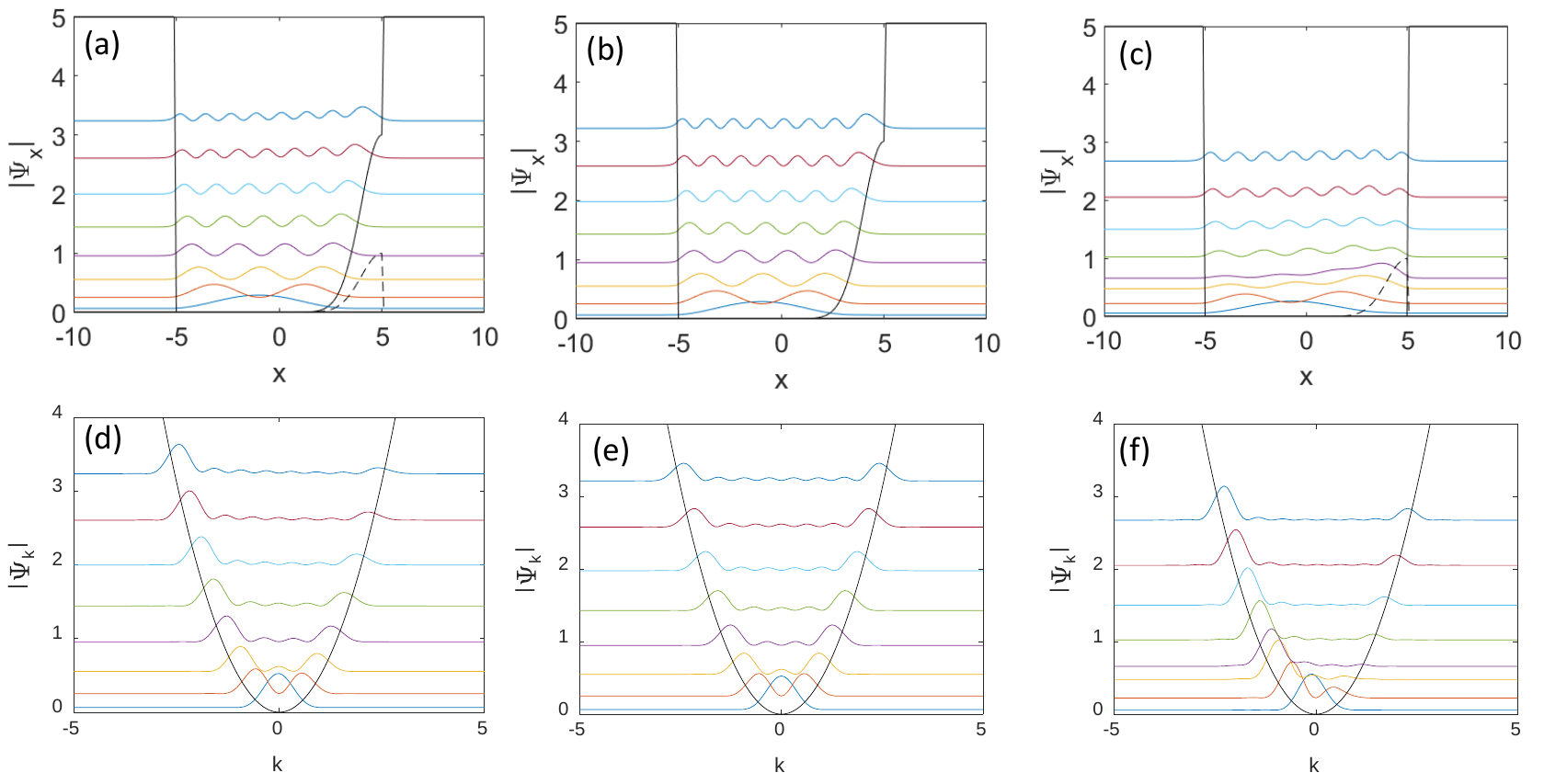}
    \caption{The 8 lowest energy eigenstates in a finite square well with height $V_0=5$ and width $a=5$ with a complex-valued Gaussian beam in real and momentum space. (a, d) showing the eigenstates with both real and imaginary Gaussian, (b, e) showing the eigenstates with only the real Gaussian and (c, f) showing the eigenstates with only the imaginary Gaussian. Here we set the heights of the real and imaginary Gaussian to be $R=3$ and $I=1$, respectively. While the width of the Gaussian is set to be $\sigma=1$.}
    \label{fig:fig1}
\end{figure*}

In the main text, we show that a imaginary Gaussian-shaped potential induced by a pump can result in the trapped states to exponentially localize in the momentum space, inducing a non-Hermitian skin effect. In this section, we present further discussion on the role of the Gaussian shape potential.

In the main text, we consider the Gaussian-shaped potential induced by the beam to be purely imaginary for simplicity. However, in general, the Gaussian can be complex-valued. Here, we consider a complex-valued Gaussian beam on the edge of the square well
\begin{equation}
    V(x)=\begin{cases}
    (R+iI)e^{-(x-x_0)^2/(2\sigma^2)},& \text{if } -a<x<a\\
    V_0,              & \text{otherwise}
\end{cases}
\end{equation}
where $R$, $I$ represent the height of the real and imaginary Gaussians, respectively, while $x_0$ is the centre of the Gaussian and $\sigma$ is the width of the Gaussian. The height of the square well is set to be $V_0=5$ while the width is $a=5$. In Fig. \ref{fig:fig1}, we plot the 8 lowest-energy eigenstates to show the effects from the real and imaginary Gaussians centred on the edge of the trap. From Fig. \ref{fig:fig1}, we can see that although a real Gaussian would shift the centre-of-mass of the eigenstates in the real space [Figs. \ref{fig:fig1}(a--b)], an imaginary Gaussian is required to induce localization of the eigenstates in momentum space [Fig. \ref{fig:fig1}(d,f)] and in the absence of the imaginary Gaussian, the eigenstates stay symmetric in the momentum space [Fig. \ref{fig:fig1}(e)].

\begin{figure*}[t]
    \centering
    \includegraphics[width=0.7\textwidth]{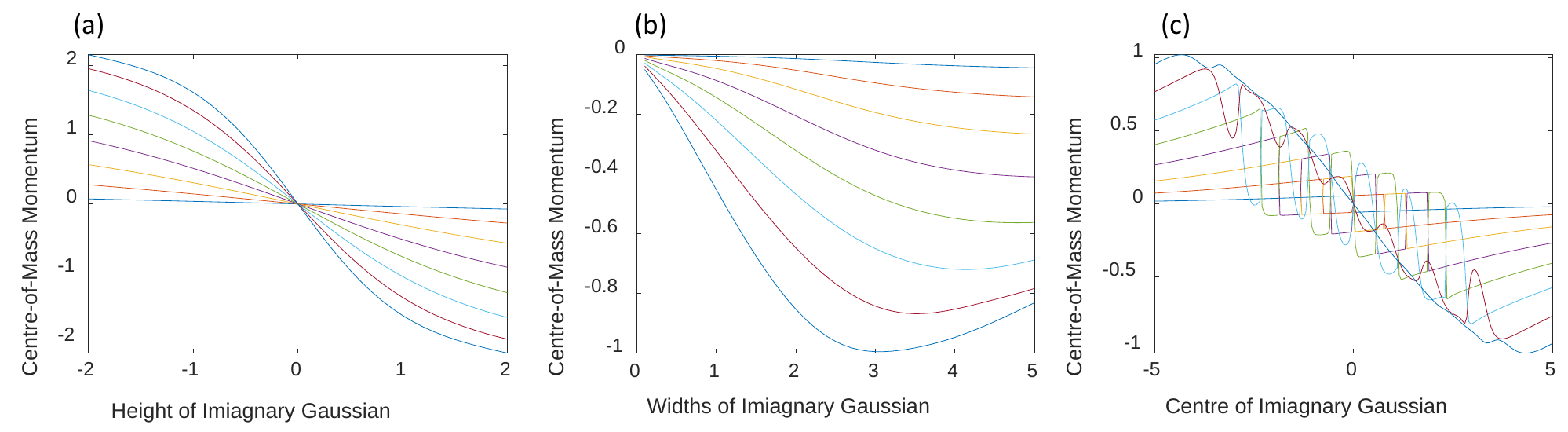}
    \caption{The centre of mass momenta of the 8 lowest energy eigenstates plotted with (a) different heights, (b) different widths and (c) different centre-of-mass position of the imaginary Gaussian. The height and width of the Gaussian are set at $I=0.5$, $\sigma=2.5$ and centred at the edge of the trap at $x_0=5$ unless specified otherwise.}
    \label{fig:fig2}
\end{figure*}

We also vary the height, the width and the centre of the imaginary Gaussian to investigate how these parameters affect the centre-of-mass momentum of the eigesntates. While from Fig. \ref{fig:fig2}(a), we can see that the strength of localization increase with higher Gaussian, the dependence on the width and the centre seem to be more complicated. The eigenstates seem to reach the strongest asymmetry at around $\sigma=3$, which seem to suggest the more anti-symetric the imaginary potential is, the stronger the localization of the eigenstateas will be [Fig. \ref{fig:fig2}(b)]. Similarly, from Fig. \ref{fig:fig2}(c), it seems that although the relation between the centre of the imaginary Gaussian and the centre-of-mass momenta of the eigenstates do not follow   simple relation, as long as the Gaussian is not symmetric (centred away from $x=0$), the eigenstates will become assymetric and localize in momentum space.

\section{Skin Effect with Two-Dimensional Potential}
\begin{figure*}[p]
    \centering
    \includegraphics[width=0.8\textwidth]{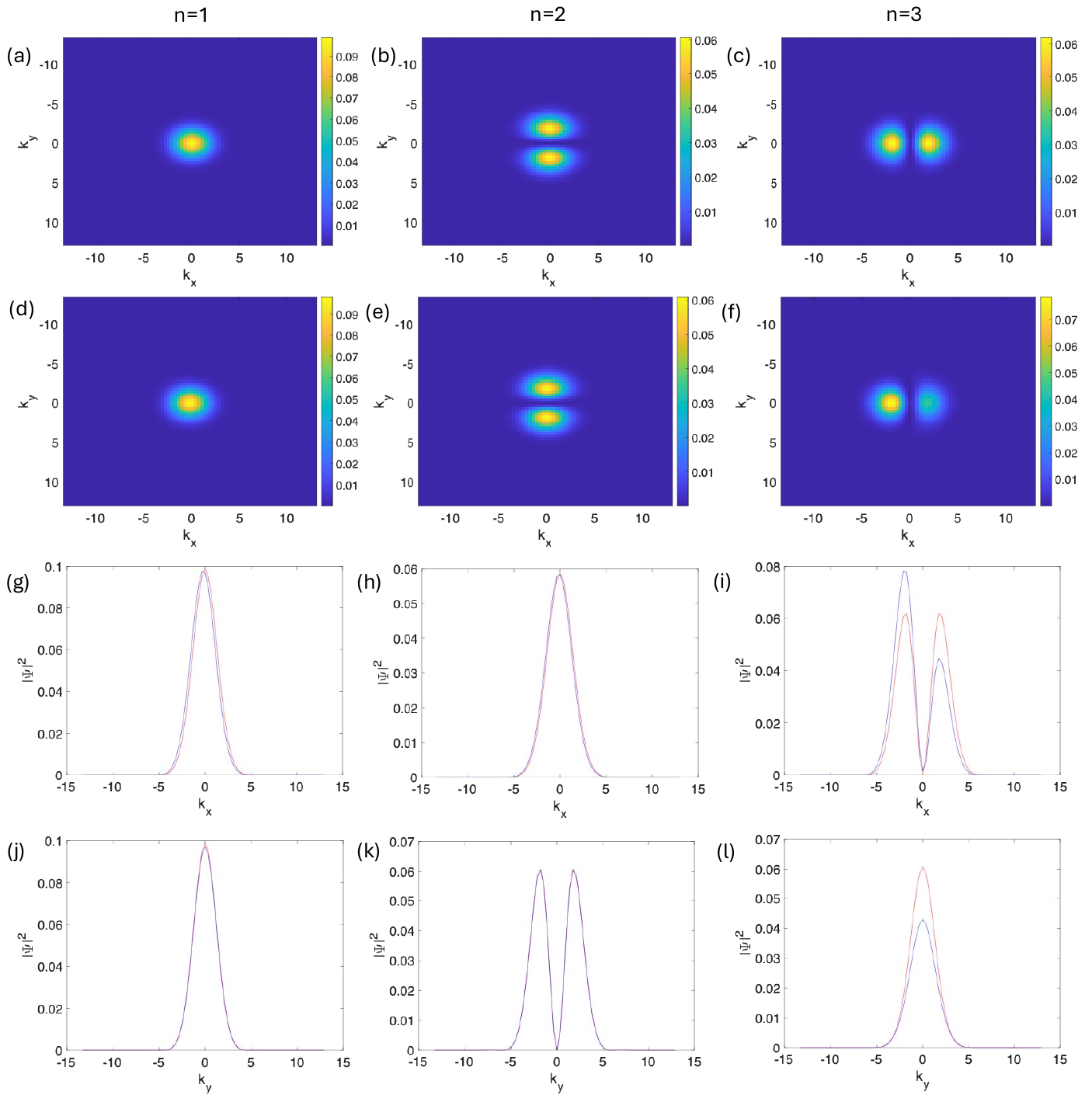}
    \caption{The distribution of eigenstates in the Hermitian (a-c) and non-Hermitian (d-f) cases. Also showing the probability distribution function along $k_y=0$ in (g,i) and $k_y\approx2$ in (h) to highlight the momentum-space localisation of the eigenstates in the $k_x$-direction in the non-Hermitian cases (blue) compared to the Hermitian ones (red), as well as the lack of localisation in the $k_y$-direction along the slices of $k_x=0$ in (j,k) and $k_x\approx 2$ in (l). Here, we set $V_0=10$, $a=1$, $R=3$, $\sigma=0.5$ and $I=2$ for the non-Hermitian case, while $I=0$ for the Hermitian case.}
    \label{fig:fig2d}
\end{figure*}

\begin{figure*}[p]
    \centering
    \includegraphics[width=0.8\textwidth]{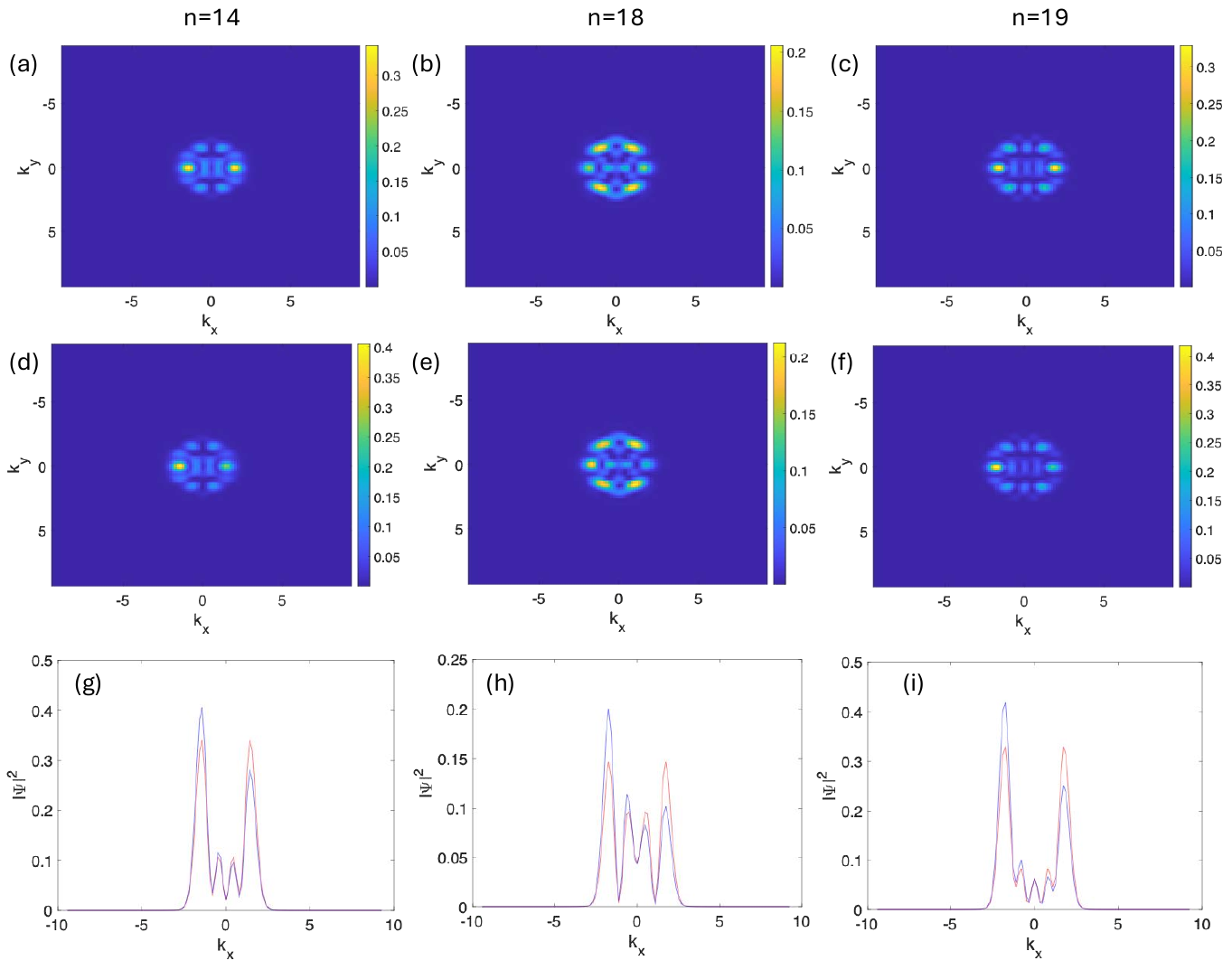}
    \caption{The distribution of higher-energy eigenstates in the Hermitian (a-c) and non-Hermitian (d-f) cases. Also showing the probability distribution function along $k_y=0$ in (g-i) to highlight the momentum-space localisation of the eigenstates in the non-Hermitian cases (blue) compared to the Hermitian ones (red). The parameters are set at $V_0=10$, $a=5$, $R=3$, $\sigma=2.5$ and $I=0.2$ for the non-Hermitian case, while $I=0$ for the Hermitian case.}
    \label{fig:fig2d2}
\end{figure*}

In the previous sections, we approximated the polaritons using an effective 1D model. Here, we will show that the momentum-space skin effect can also be induced in a two-dimensional system and the numerical results are presented in Figs.~\ref{fig:fig2d} and ~\ref{fig:fig2d2}.

We consider a two-dimensional non-Hermitian system with a complex-valued potential
\begin{equation}
    \begin{split}
        V(x)&=V_{trap}(x)+V_{pump}(x)\\
        V_{trap}(x)&=V_0\bigg(1-\Big(e^{-\frac{x^2+y^2}{a^2}}\Big)^{20}\bigg)\\
        V_{pump}(x)&=(R+iI)e^{-\frac{(x-\delta_x)^2}{2\sigma^2}-\frac{(y-\delta_y)^2}{2\sigma^2}}.
    \end{split}
\end{equation}
Here, $V_{trap}$ denotes a 2D trapping potential arising from the sample structure approximated using an inverted super Gaussian with a height of $V_0=10$ and a radius of $a/2$. $V_{pump}$ represents a complex-valued Gaussian-shaped potential induced by the pump, where the real part has a height of $R$, the imaginary part has a height of $I$, and both Gaussians have width of $\sigma$ and are centred at $(\delta_x,\delta_y)$. In this configuration, the height $I$ is the main non-Hermitian parameter corresponding to the parameter $V_i$ in the main text.

In Fig.~\ref{fig:fig2d}, we plot the numerical simulation of the low-energy eigenstates in the absence (when $I=0$) [Figs.~\ref{fig:fig2d} (a-c)] and the presence [Figs.~\ref{fig:fig2d} (d-f)] of the imaginary-valued Gaussian pump. The pump-induced potential is centred at the edge of the trap at $\delta_x=a$, $\delta_y=0$. Note that the real part of the pump-induced potential is included in the $I=0$ case.

We show that in this configuration the momentum-space wavefunction localise towards the negative $k_x$-direction. We highlighted the localisation by directly comparing the Hermitian and non-Hermitian cases in Figs.~\ref{fig:fig2d} (g-i) along the cross-section at $k_y=0$ for $n=1,3$ and at $k_y\approx2$ for $n=2$. Note that states $n=1,3$ are observed in the experiment while the $n=2$ state is missed since it has a node at $k_y=0$. However, the profiles along the $k_x$ direction clearly shows that all three eigenstates are localised. Similarly, in Figs.~\ref{fig:fig2d} (j-l) we  show the cross-section along $k_y$ at $k_x=0$ for $n=1,2$ and at $k_x\approx2$ for $n=3$ to show the absnece of localisation in the $k_y$ direction. This shows that momentum-space localisation induced by the complex-valued pump in the 2D system works very similarly to the quasi-1D cases discussed in the previous sections and in the main text.

To match the experimental results presented in Figs.~2 and 3 of the main text, we also simulated the case with a wider trap at $a=5$, which allows more trapped eigenstates. The numerical results are presented in Fig.~\ref{fig:fig2d2} in a similar manner as Figs.~\ref{fig:fig2d}. The comparison of the momentum-space profiles [see Fig.~\ref{fig:fig2d2}(g-i)] shows that the localisation also occurs in higher energy states similarly to the experimental results in Figs. 2 and 3 in the main text.

\section{Experimental setup} 

\begin{figure}[h]
    \centering
    \includegraphics{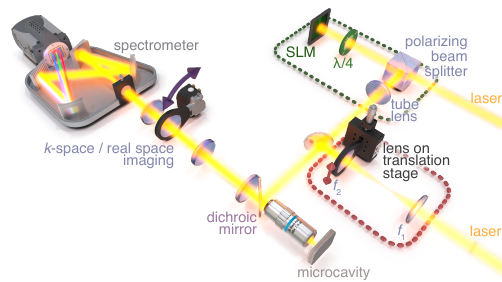}
    \caption{Scheme of the experimental setup.}
    \label{fig:Setup}
\end{figure}

The scheme of the experimental setup is presented in Fig.\,\ref{fig:Setup}. The setup allowed to control the asymmetry of the excitation profile by two separate methods.

The first method used a spatial light modulator (SLM). The laser light was reflected by a polarizing beam splitter. After passing through a quarter-wave plate oriented at a 45$^\circ$ angle, the polarization of the laser beam is changed to circular. Then, the polarization of the beam can be modified by a reflective SLM. After the modulation, the beam passes through the quarter-wave plate and the polarizing beam splitter. The pattern sent to the SLM controls the intensity of light that passes through the polarizing beam splitter, what can be used to create asymmetric beam profiles. The image of the SLM chip is then projected by a tube lens and a microscope objective onto the sample mounted in a cryostat. 

\begin{figure*}[h]
    \centering
    \includegraphics{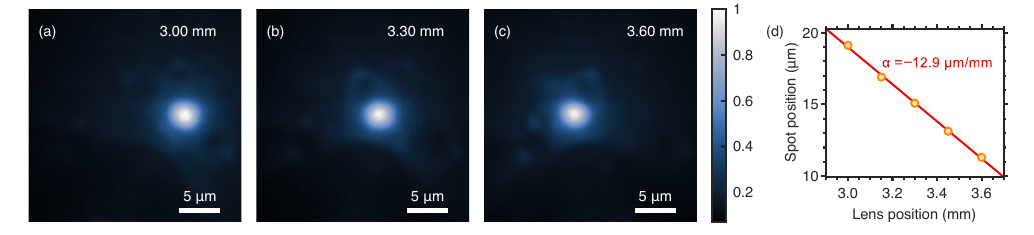}
    \caption{Calibration of the excitation spot position versus vertical shift of the lens. (a--c) Real space emission profiles at three different positions of the $f_2$ lens: 3.00\,mm, (a), 3.30\,mm (b) and 3.60\,mm (c). (d) Position of the emission spots at different shifts of the lens. Red line shows linear fit to the experimental points.  }
    \label{fig:VerticalCalibration}
\end{figure*}

Another method of achieving asymmetric optical pumping is based on translation of a lens. The parallel laser beam is expanded by a telescope made of two convex lenses with focal lengths: $f_1 = 125$\,mm and $f_2 = 300$\,mm. The second lens is mounted on a motorized vertical translation stage. The translation of the lens (in the range of approx. $\pm 0.3$\,mm) allows for a precise and repeatable shift of the Gaussian excitation spot vertically across the sample. The calibration of this vertical shift is presented in Fig.\,\ref{fig:VerticalCalibration}. Fig.\,\ref{fig:VerticalCalibration}(a--c) present real space emission profiles from a planar (unstructured) part of the sample at three different positions of the $f_2$ lens. The position of the emission spot plotted vs. location of the lens is shown in Fig.\,\ref{fig:VerticalCalibration}(d). The linear fit providing the slope of $\alpha = -12.9$\,$\upmu$m$/$mm can then be used to relate the position of the excitation spot on the sample to the location of the lens.

Independently of the excitation beam shaping technique, the laser beam is reflected from a dichroic mirror and focused on the sample with a microscope objective. The light emitted from the sample is collected by the same microscope objective in a reflection geometry. A set of convex lenses on the detection path allow to project either real space image of the sample or the back focal plane of the objective onto the entrance slit of a spectrometer. The back focal plane of the objective optically realizes the Fourier transform of the emitted signal, used here to obtain angle-resolved spectra \cite{Cueff2024}.  

All measurements were performed at temperature of 4.5\,K. For the excitation Ti:sapphire CW laser was used, tuned to 723\,nm wavelength (1.715\,eV). The excitation laser was modulated with an AOM with a duty cycle of $0.5\%$. 

\section{Extended experimental data}
\begin{figure}[h]
    \centering
    \includegraphics{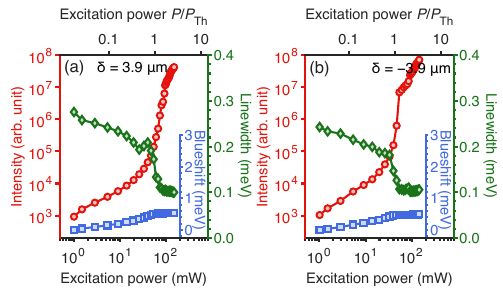}
    \caption{Extended data for Fig.\,3. (a,b) Input-output curve, blueshift, and linewidth of the condensing mode for $\delta = 3.9$\,$\upmu$m (a) and $\delta = -3.9$\,$\upmu$m (b).}
    \label{fig:fig3pow}
\end{figure}

\begin{figure}[h]
    \centering
    \includegraphics{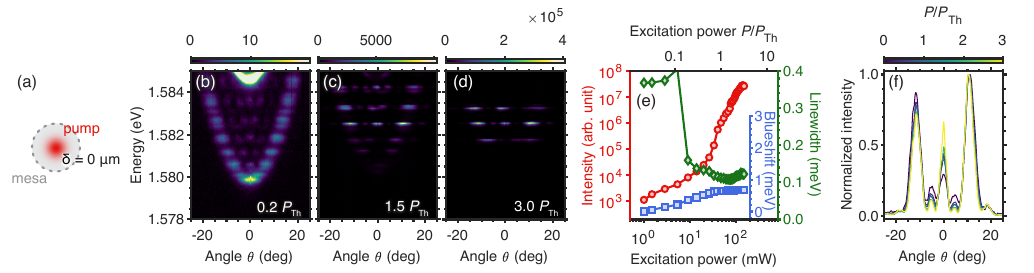}
    \caption{Extended data for Fig.\,3 corresponding to symmetric excitation ($\delta = 0$\,$\upmu$m)  (a) Schematic of pump configuration centered with a 10\,$\upmu$m wide mesa trap. (b--d) Angle-resolved PL below to above threshold. (e) Input-output curve, blueshift, and linewidth of the condensing mode. (f) Normalized (with respect to maximum) momentum-space profile of the mode as a function of pump power. }
    \label{fig:fig3ext}
\end{figure}

\begin{figure}[h]
    \centering
    \includegraphics{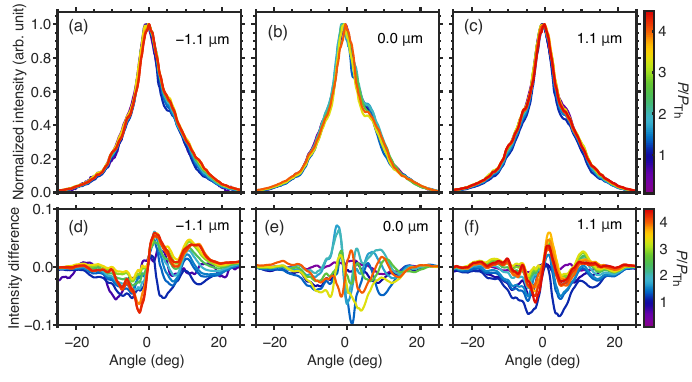}
    \caption{Extended data for Fig.\,4. (a,b,c) Intensity-normalized angle-resolved photoluminescence for three pumping configurations. (d,e,f) Corresponding difference from the emission profile at the lowest excitation power. }
    \label{fig:fig4ext}
\end{figure}

\begin{figure}[H]
    \centering
    \includegraphics[scale=1]{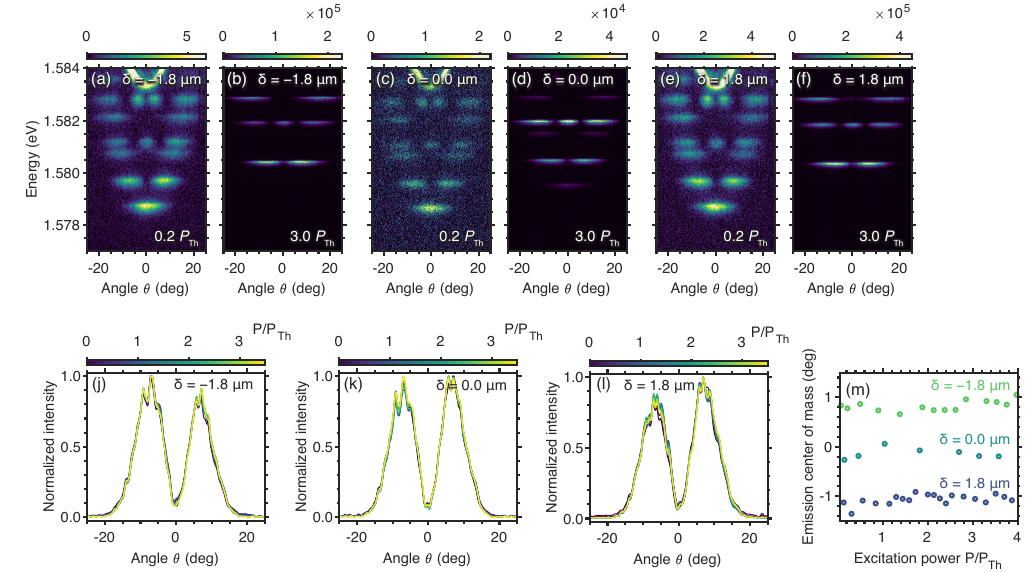}
    \caption{Condensation in the excited state of a 5\,$\upmu$m wide pillar. (a--f) Angle-resolved emission spectra for the laser excitation spot displacements $\delta = -1.8$\,$\upmu$m, $0.0$\,$\upmu$m, $1.8$\,$\upmu$m and excitation powers of $0.2$\,$P_\textrm{Th}$ and $3.0$\,$P_\textrm{Th}$. (g--i) Emission intensity, spectral linewidth and energy blueshift as a function of excitation power. (j--l) Angle-resolved emission profile of the first excited state in the pillar. (m) Emission angle center of mass for increasing excitation power.}
    \label{fig:fig7ext}
\end{figure}

\begin{figure}[H]
    \centering
    \includegraphics[scale=1]{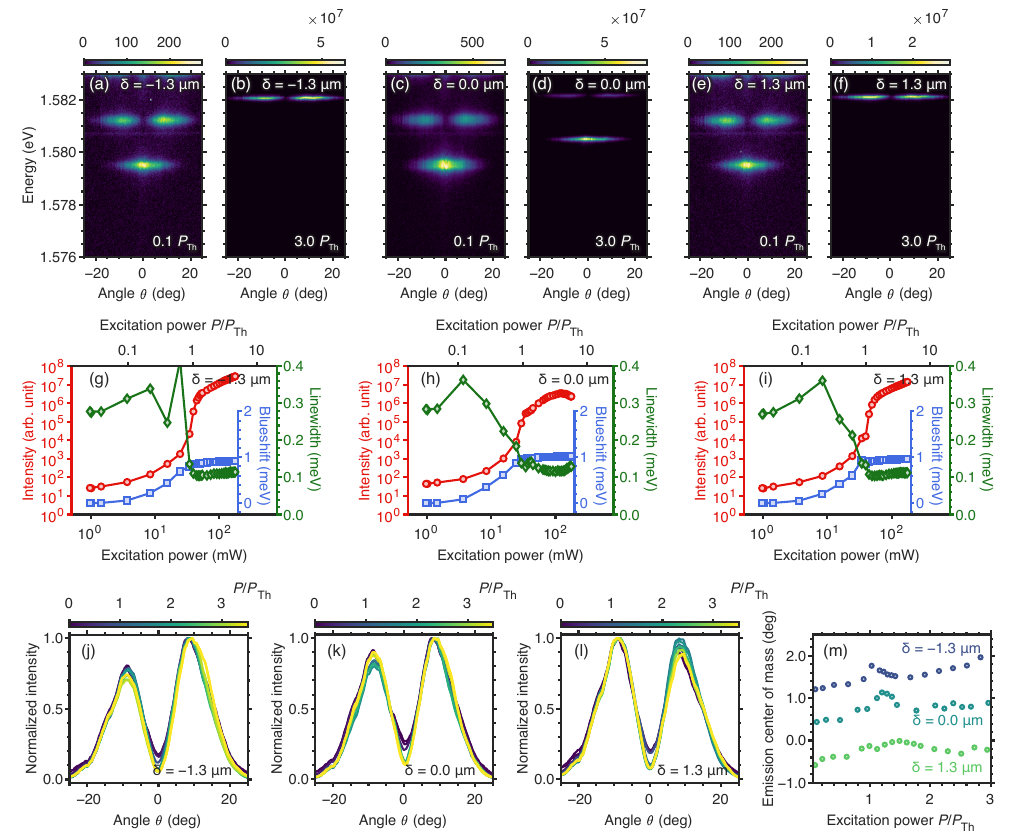}
    \caption{Condensation in the excited state of a 3.5\,$\upmu$m wide pillar. (a--f) Angle-resolved emission spectra for the laser excitation spot displacements $\delta = -1.3$\,$\upmu$m, $0.0$\,$\upmu$m, $1.3$\,$\upmu$m and excitation powers of $0.1$\,$P_\textrm{Th}$ and $3.0$\,$P_\textrm{Th}$. (g--i) Emission intensity, spectral linewidth and energy blueshift as a function of excitation power. (j--l) Angle-resolved emission profile of the first excited state in the pillar. (m) Emission angle center of mass for increasing excitation power.}
    \label{fig:fig6ext}
\end{figure}

Figure~\ref{fig:fig3pow} show the power dependence of the intensity, blueshift, and linewidth corresponding to the experimental conditions in Fig.~3 of the main text. 
Figures\,\ref{fig:fig3ext} and \ref{fig:fig4ext} present data corresponding to the symmetric pump, not shown directly in the main text. Two sets of extended data are also provided for the case of a smaller traps ($D=3.5$~$\mu$m) in Fig.~\ref{fig:fig6ext} and ($D=3.5$~$\mu$m) in Fig.~\ref{fig:fig7ext}. These experiments are very similar to the power series measurements presented in Fig.~3 of the main text.

\section{Determination of center-of-mass momentum uncertainty}

\begin{figure}[H]
    \centering
    \includegraphics{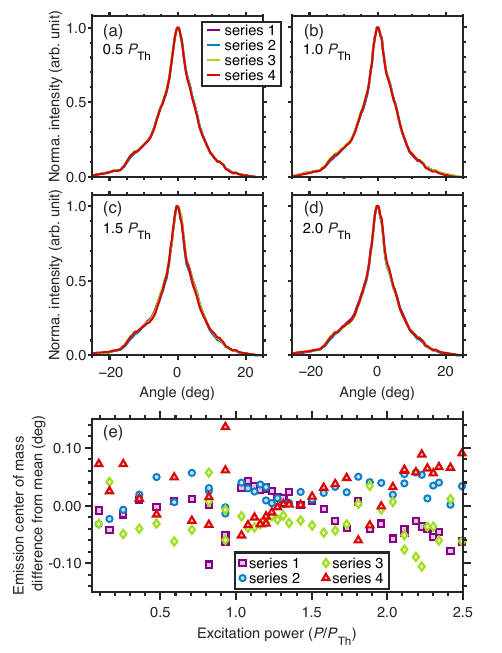}
    \caption{Uncertainty of angle-resolved imaging. (a--d) Angle-resolved emission spectra collected at four independent measurement series at four different excitation powers: $0.5$\,$P_\textrm{Th}$ (a), $1.0$\,$P_\textrm{Th}$ (b), $1.5$\,$P_\textrm{Th}$ (c) and $2.0$\,$P_\textrm{Th}$ (d). (e) Center of mass emission angle for increasing excitation power in four independent measurement series.}
    \label{fig:fig5ext}
\end{figure}

To estimate the experimental uncertainty of the measured emission angle center of mass, we performed four independent measurement series collecting angle-resolved PL spectra for increasing pump powers. The data collected from a $2$\,$\upmu$m wide pillar is presented in Fig.\,\ref{fig:fig5ext}. Figure\,\ref{fig:fig5ext}(a--d) presents spectra collected at: $0.5$\,$P_\textrm{Th}$, $1.0$\,$P_\textrm{Th}$, $1.5$\,$P_\textrm{Th}$ and $2.0$\,$P_\textrm{Th}$, respectively. 
The obtained emission angle center of mass is presented in Fig.\,\ref{fig:fig5ext}(e). The deviations form the mean do not visibly depend on the excitation power. The standard deviation of 0.044\,deg was used in Fig.\,4 in the main text as an uncertainty in the experimental determination of the emission center of mass.

\section{Experimental estimation of the momentum space exponential decay constant $\xi_n$}

The estimation of the momentum space decay constant $\xi$ (or inverse localisation length) from the experimental data was performed based on Eq.~(5) from the main manuscript:
\begin{equation}
    \left|\psi_n(p;V_i)\right|^2 = e^{-2\xi_n p/\hbar} \left|\psi_n(p;0)\right|^2.
    \label{eq:localisation}
\end{equation}
However, here, for simplicity, we use the measured distribution for symmetric excitation (when $\delta=0$) as the reference $\psi_n(p;0)$. Hence, the equation above represents the approximate connection between the wavefunction corresponding to the asymmetric excitation $\psi_n(p;\delta)$ with the wavefunction for the symmetric case $\psi_n(p;\delta=0)$.

The value of $\xi_n$ was directly fitted to the experimental emission intensity profiles $\left|\psi_n(p)\right|^2$ using Eq.~\eqref{eq:localisation}. In the fitting, the probability distribution corresponding to the symmetric case $\left|\psi_n(p;\delta=0)\right|^2$ was estimated as the asymmetric distribution averaged over the absolute value of momentum $p$ $\left( \left|\psi_n(p;\delta=0)\right|^2 \approx \left|\psi_n(|p|)\right|^2 \right)$.

\begin{figure*}[ht]
\centering
\includegraphics{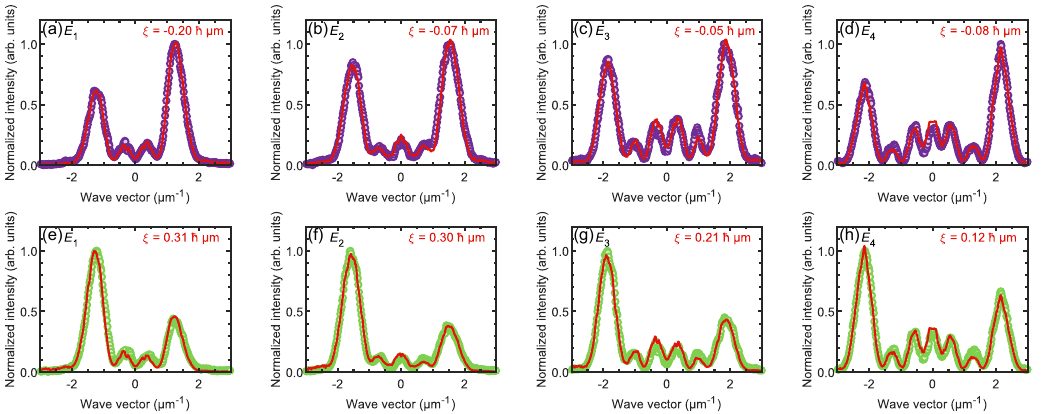}             
\caption{ Momentum space emission profiles from Fig.~2 of the main manuscript of four selected energy states ($E_1 = 1.5800$\,eV, $E_2 = 1.5808$\,eV, $E_3 = 1.5816$\,eV, $E_4 = 1.5824$\,eV) for (a--d) $\delta<0$ and (e--h) $\delta>0$. The experimental results (coloured circles) are fitted (red lines) to extract momentum space localisation length $\xi$.
}%
\label{im:ExperimentalFig1} 
\end{figure*}

\begin{figure*}[ht]
\centering
\includegraphics{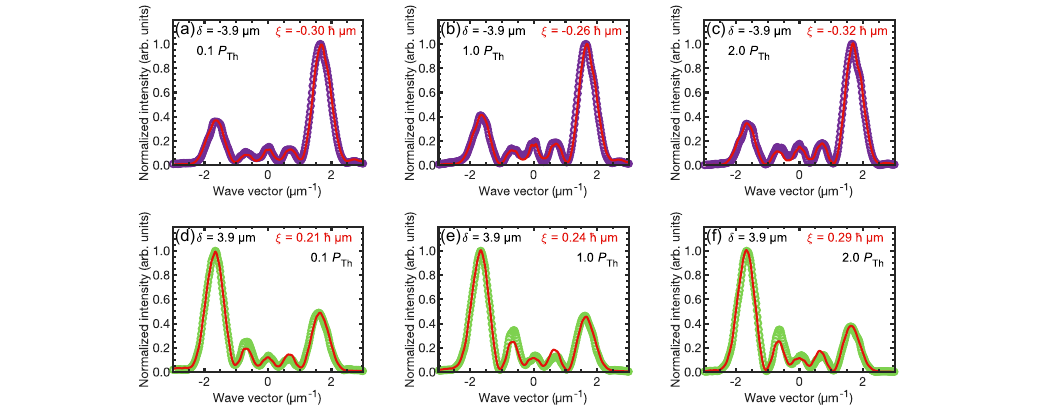}             
\caption{  Momentum space emission profiles from Fig.~3 of the main manuscript for increasing excitation power (0.1\,$P_\text{Th}$, 1.0\,$P_\text{Th}$, 2.0\,$P_\text{Th}$) at (a--c) $\delta = -3.9$\,$\upmu$m and (d--f) $\delta = 3.9$\,$\upmu$m. The experimental results (coloured circles) are fitted (red lines) to extract momentum space localisation length $\xi$.}%
\label{im:ExperimentalFig2} 
\end{figure*}

The results of the fitting are presented in Figs.\,\ref{im:ExperimentalFig1} and \ref{im:ExperimentalFig2}. Fig.\,\ref{im:ExperimentalFig1} shows data corresponding to the several excited states of the 10\,$\upmu$m-wide trap corresponding to Fig.\,2 of the main manuscript. Fig.\,\ref{im:ExperimentalFig2} presents the fitting  excitation power dependent series corresponding to Fig.\,3 of the main manuscript.

\section{Numerical model}

\begin{figure*}
    \centering
    \includegraphics{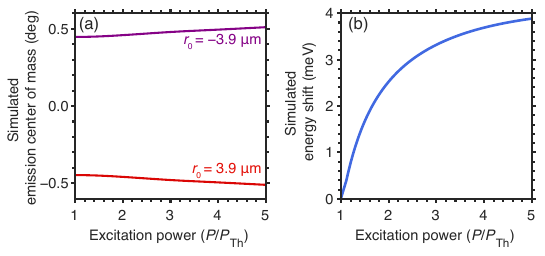}
    \caption{Simulated d-like condensate in a 10\,$\upmu$m wide trap with asymmetric pump. (a) Center of mass momentum and (b)\,condensate energy blueshift as a function of pump power.}
    \label{fig:Theory10um}
\end{figure*}

\begin{figure*}
    \centering
    \includegraphics{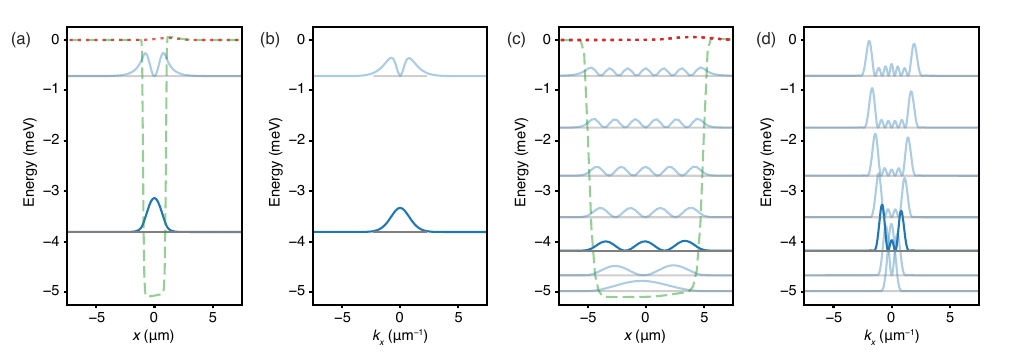}
    \caption{Simulated condensate profiles. (a,b) Condensate profiles in real (a) and momentum space (b) in a 2\,$\upmu$m wide pillar. (c,d) Condensate profiles in real (c) and momentum space (d) in a 10\,$\upmu$m wide pillar.}
    \label{fig:TheoryProfiles}
\end{figure*}

For the numerical modeling, we employ the generalized Gross-Pitaevskii equation with reservoir coupling
$$
\begin{aligned}
&i\hbar\dfrac{\partial \psi(\mathbf{r},t)}{\partial t}  +  \dfrac{\hbar^{2}}{2m}\nabla^2\psi(\mathbf{r},t)-U(\mathbf{r})\psi(\mathbf{r},t)-G_{C}|\psi(\mathbf{r},t)|^2\psi(\mathbf{r},t) \\&-G_{R} n_{R}(\mathbf{r})\psi(\mathbf{r},t) -\dfrac{i\hbar}{2}\bigl(R n_{R}(\mathbf{r})-{\Gamma}_{C}\bigr)\psi(\mathbf{r},t)=0, \\
&\frac{\partial n_R(\mathbf{r},t)}{\partial t} =  -\left(\Gamma_R +R |\psi(\mathbf{r},t)|^2\right) n_R(\mathbf{r},t)(\boldsymbol{r}, t)+P(\mathbf{r})\:,
\end{aligned}
$$
where $\psi(\mathbf{r},t)$ is a collective wave function of polaritons, 
and $n_R(\mathbf{r},t)$ is an exciton reservoir density. 
We consider a one-dimensional (1D) system where $\mathbf{r} \equiv x$, analyzing both non-stationary (time-dependent) and stationary regimes. To recast the dynamical equations into a time-independent problem, the solution is sought in the steady-state form:
\begin{equation}
\psi(\mathbf{r},t) = \psi(\mathbf{r})\, e^{-i E t} , \quad  
n_R(\mathbf{r},t) = n_R(\mathbf{r}),
\end{equation}
where $\psi(\mathbf{r})$ and $n_R(\mathbf{r})$ are localized spatial profiles satisfying $|\psi(\mathbf{r})| \to 0$ and $n_R(\mathbf{r}) \to 0$ as $|\mathbf{r}| \to \infty$.

By neglecting the nonlinearity of the condensate and retaining only the coupling with the reservoir, the system simplifies to a linear eigenvalue problem. Because the system inherently includes both gain and dissipation, these linear eigenvalues are complex-valued. Conversely, retaining the condensate's self-interaction terms yields the full nonlinear problem. In solving the full nonlinear problem, we focus on solutions that are neither decaying nor growing over time, restricting our interest to purely real energies $E$. This real energy $E$ must be determined self-consistently alongside the spatial profile of the field.

The governing equations are solved numerically by discretizing the continuous spatial domain into a computational grid. To approximate the spatial second derivative, a high-accuracy 8th-order finite difference scheme is employed, reducing the differential equation to a matrix equation. We implement an iterative solver that utilizes the solutions of the linear eigenvalue problem as a primary initial approximation. While this high-order scheme yields excellent numerical convergence and high fidelity, its convergence region is limited. Consequently, the solver is highly sensitive to the initial guess, making the an accurate initial state provided by the linear problem essential for successful convergence. Due to the global phase invariance of the system, we remove this degree of freedom by fixing the phase to zero at a reference point. Setting this phase condition provides the additional algebraic constraint required to uniquely determine the real energy $E$ during the root-finding process. Finally, the stability of these solutions is confirmed by dynamically evolving the system from random noise over a range of powers along the slope of the condensate energy curve up to saturation.

We use representative 
parameters: the effective mass $m =36 \cdot 10^{-6} m_{\mathrm{e}}$ related to the free electron mass $m_{\mathrm{e}}$, and the trapping potential taken in the form of super Gauss 
\begin{equation*}
U(\mathbf{r}) = U_0 \left[ 1 - \exp\left( -\frac{|\mathbf{r}|^{2q}}{{(0.5\Sigma_U)}^{2q}}\right) \right]
\end{equation*} with the potential depth $U_0 = 5.1~\mathrm{meV}$, order $q=9$ and mesa diameters $ \Sigma_U =2$~$\upmu$m (narrow pillar, Fig.~4), $\Sigma_U =10$~$\upmu$m (wide pillar, Fig.~\ref{fig:Theory10um}). 
The stimulated scattering from the reservoir to polaritons is $R$. The strengths of polariton-polariton and polariton-reservoir interactions are given by the expressions 
$G_C = \alpha |X|^4$ and 
$G_R = 2 \alpha  |X|^2$, where
$\alpha = 2.557 \times 10^{-18} \ \text{eV}\cdot\text{m}^2$ and excitonic Hopfield coefficient
$|X|^2 = 0.35$. The constants $\Gamma_C = 6.2 \times 10^{10} \ \text{s}^{-1} $ and $\Gamma_R = 0.8 \times 10^{10} \ \text{s}^{-1}$ stand for the decay rates of condensed polaritons and reservoir, respectively. These losses are compensated by an external off-resonant time-independent optical pump with the rate $P(\boldsymbol{r})$ of profile 
\begin{equation*}
P(\mathbf{r}) = P_0 \exp\left( -\frac{|\mathbf{r} - \mathbf{r}_0|^2}{(0.5\Sigma_P)^2} \right)
\end{equation*}
shifted by $r_0$ with respect to the center of a pillar. The power is swept by varying the coefficient $ C_{\text{P}} $ in its amplitude scaled as 
$P_0 = C_{\text{P}} \frac{\Gamma_C \Gamma_R}{R} $, where $R = 3.58 \times 10^{-3}~\text{m}^2/\text{s} $.
For the narrow and wide pillars, the pump widths and shifts are set to $\Sigma_P = 2.5$~$\upmu$m, $r_0 = \pm 1.1$~$\upmu$m  and $\Sigma_P = \sqrt{2} \cdot 3.2$~$\upmu$m, $r_0 = \pm 3.9$~$\upmu$m. The resulting profiles in real and momentum space near the power threshold [as defined in the main text, Figs.~4(f,h), and in Fig.~\ref{fig:Theory10um}] for positive $r_0$ are presented in Fig.\,\ref{fig:TheoryProfiles}.

\clearpage
\bibliography{ref_polariton}